\documentclass[prd,showpacs,twocolumn,superscriptaddress,floatfix,nofootinbib,10pt]{revtex4-2}
\usepackage{setspace}
\usepackage[utf8]{inputenc}
\usepackage{float}
\usepackage{amsmath,amssymb,amsfonts,bm}
\usepackage{graphicx}
\usepackage{multirow}
\usepackage{lipsum}
\usepackage[usenames,dvipsnames]{color}
\allowdisplaybreaks
\usepackage[
colorlinks=true,
linkcolor=blue,
breaklinks=true,
urlcolor=blue,
citecolor=blue]{hyperref}

\usepackage{orcidlink}
\usepackage{subfigure}
\usepackage{soul}
\usepackage{xcolor} 
\usepackage{physics}
\usepackage{booktabs}  
\usepackage{ulem} 
\usepackage{tikz-feynman}
\usepackage{makecell}
\definecolor{green}{rgb}{0,0.6,0}
\newcommand{\mev}{\textrm{ MeV}}
\newcommand{\kev}{\textrm{ keV}}

\newcommand{\GXNU}{\affiliation{Department of Physics, Guangxi Normal University, Guilin 541004, China}}

\newcommand{\GXZD}{\affiliation{Guangxi Key Laboratory of Nuclear Physics and Technology, Guangxi Normal University, Guilin 541004, China}}

\newcommand{\ZZU}{\affiliation{School of Physics, Zhengzhou University, Zhengzhou 450001, China}}

\newcommand{\IFIC}{\affiliation{Departamento de F\'{\i}sica Te\'orica and IFIC, Centro Mixto Universidad de
		Valencia-CSIC Institutos de Investigaci\'on de Paterna, Apartado 22085,
		46071 Valencia, Spain}}

\begin{document}
	\title{The $D_{s0}^*(2317)^+$ decay to $D_s^+\pi^0$ and $D_s^{*+}\gamma$}
	
	\begin{abstract}
	We study the strong decay of $D_{s0}^*(2317)^+$ to $D_s^+ \pi^0$ considering the coupled channels of $D^0 K^+, D^+ K^0, D_s^+ \eta$ and $D_s^+ \pi^0$ within a picture for the interaction based on the local hidden gauge approach. We also address the problem of the radiative decay to $D_s^{*+} \gamma$, using the same information obtained from the molecular picture. We obtain a strong width of the $D_{s0}^*(2317)^+$ of about $77 \,\rm keV$ from coupled channels interaction and a radiative decay of about $1.7 \,\rm keV$. We also show that the extra consideration of $\pi^0-\eta$ mixing can double the strong decay width to values around $140 \,\rm keV$. The anomalous terms for the radiative decay are considered for the first time, but they are found negligible. We make a thorough discussion of this and other results to the light of the recent measurement of Belle for the ratio of these two decay modes, and make a call for the precise measurement of the two decay widths independently to clarify the present situation concerning the nature of the $D_{s0}^*(2317)^+$ state.
	\end{abstract}
	
	\author{Pei-Sen Su\orcidlink{0009-0000-1302-8123}}%
	\GXNU%
	\GXZD%
	\author{}%
	\author{Wen-Tao Lyu\orcidlink{0009-0007-8470-2761}}%
	\ZZU%
	\author{Wei-Hong Liang\orcidlink{0000-0001-5847-2498}}%
	\email{liangwh@gxnu.edu.cn}
	\GXNU%
	\GXZD%
	\author{Eulogio Oset\orcidlink{0000-0002-4462-7919}}%
	\email{Oset@ific.uv.es}
	\GXNU%
	\IFIC%

\maketitle

\section{Introduction}\label{sec:Intr}
The radiative decay of resonances has been often advocated as a good test to learn about their nature~\cite{Wang:2006mf,Kalashnikova:2005zz}. 
In particular the $D_{s0}^*(2317)^+ \to D_s^{*+}\gamma$ decay has been thoroughly studied from different points of view, quark models, vector meson dominance and QCD sum rules, with varied results~\cite{Godfrey:2003kg,Colangelo:2003vg,Bardeen:2003kt,Fayyazuddin:2003aa,Ishida:2003gu,Azimov:2004xk,Colangelo:2005hv,Close:2005se,Liu:2006jx,Wang:2006zw}. 
On the other hand, mounting evidence has been piling up about the molecular nature of the $D_{s0}^{*}(2317)$ as a result of the coupled channel interaction with the $D^{0}K^{+}, D^{+}K^{0}, D_{s}^{+}\eta$ channels~\cite{vanBeveren:2003kd,Barnes:2003dj,Chen:2004dy,Kolomeitsev:2003ac,Gamermann:2006nm,Guo:2006rp,Yang:2021tvc,Liu:2022dmm}.
Their claims are also supported by recent lattice QCD calculations~\cite{Mohler:2013rwa,Lang:2014yfa,Bali:2017pdv,Cheung:2020mql}.  
A detailed analysis of lattice data in Ref.~\cite{MartinezTorres:2014kpc} supports this picture and finds around $72\%$ probability for the $DK$ component in the $D_{s0}^{*}(2317)$ state.

From the molecular perspective there are also early works evaluating the
$D_{s0}^*(2317) \to D_s^* \gamma$ decay~\cite{Faessler:2007gv,Gamermann:2007bm,Lutz:2007sk} and more recently in Ref.~\cite{Cleven:2014oka} and the update of this latter work in Ref.~\cite{Fu:2021wde}.

The reason to reopen the issue is the recent work reported by the Belle collaboration~\cite{Belle-II:2025dzk}, where the branching fraction ratio of the radiative decay to the strong decay of the $D_{s0}^*(2317)$ is measured for the first time with the result~\cite{Belle-II:2025dzk}
\begin{equation}
	\dfrac{\mathcal{B}(D_{s0}^*(2317)^+ \to D_s^{*+} \gamma)}{\mathcal{B}(D_{s0}^*(2317)^+\to D_s^+\pi^0)}=(7.14\pm 0.70\pm0.23) \times 10^{-2}.
\end{equation}
This result is at variance with present results reported in the PDG~\cite{ParticleDataGroup:2006fqo} that say this ratio is smaller than $0.059$.

The new measurement also gives us an opportunity to update the result of Ref.~\cite{Gamermann:2007bm} to the light of recent advances in the field. 
The molecular state of $DK, D_s \eta$ is now tackled with an interaction based on the extention of the local hidden gauge approach~\cite{Bando:1984ej,Bando:1987br,Meissner:1987ge,Nagahiro:2008cv} to the charm sector, as done in Ref.~\cite{Ikeno:2023ojl}, which proceeds by the exchange of vector mesons. 
A new method is also used to evaluate the loops in the radiative decay, tied to the findings of the strong interaction in Ref.~\cite{Ikeno:2023ojl}, and the contribution of loops involving anomalous terms is also considered.

On the other hand, the measurement of Ref.~\cite{Belle-II:2025dzk} involves the ratio of the radiative decay to the strong one of the $D_{s0}^*(2317)$. 
Then we also tackle the problem of the strong decay of the $D_{s0}^*(2317)$ to the isospin forbidden mode $D_s\pi^0$, which, according to BESIII results~\cite{BESIII:2017vdm}, basically accounts for all the $D_{s0}^*(2317)$ decay width. 
We evaluate this decay by implementing a coupled channel calculation with the $D^0 K^+, D^+ K^0, D_s^+\eta, D_s^+ \pi^{0}$ channels, and also using triangle diagrams to see the consistency of the approach.

The issue of the strong decay of the $D_{s0}^*(2317)$ has also received attention in the literature. 
In Ref.~\cite{Guo:2006fu}, a small decay width of $8.7 \kev$ was obtained from the consideration of $\eta-\pi^{0}$ mixing. 
But in Ref.~\cite{Faessler:2007gv} it was shown that the consideration of the different mass of neutral and charged kaons was important and resulted in much bigger widths. 
In addition, electromagnetic corrections were considered in Ref.~\cite{Guo:2008gp}, and the results were updated in Ref.~\cite{Liu:2012zya}. 
Further work was done in Ref.~\cite{Fajfer:2015zma} and Ref.~\cite{Guo:2018kno}. 
From inspection of all these works one can see that there is a large span in the predictions for the decay width, from $2.4\kev$ to $180 \kev$.
More recently a work considering coupled channels $D^+ K^0, D^0 K^+, D_s^+ \eta$ and $D_s^+ \pi^0$ has been conducted~\cite{Achasov:2025hkv} and within uncertainties of the model, results for the width in line with those in Ref.~\cite{Liu:2012zya} are obtained. 
A different approach is followed in Ref.~\cite{Yue:2025wcl}, where information from the $T_{c\bar{s}0}^{a}(2327)$ state is used to find a bracket of values $[63-209] \kev$ for the strong decay width. 
It is worth having a fresh look of the problem, and since the experimental work of Ref.~\cite{Belle-II:2025dzk} only provides the ratio of the radiative width to the strong one, the consideration of the two decay modes with the same framework, as we do here, becomes advisable.

\section{formalism}\label{sec:form}
\subsection{Strong decay $D_{s0}^{*}(2317)^{+}\to D_{s}^{+}\pi^{0}$}

We tackle first this problem and follow the approach of Ref.~\cite{Ikeno:2023ojl}, using the local hidden gauge approach for the interaction with the exchange of vector mesons \cite{Bando:1984ej,Bando:1987br,Meissner:1987ge,Nagahiro:2008cv}. 
In Ref.~\cite{Ikeno:2023ojl}, the isospin conserving channels $D^0 K^+, D^+ K^0, D_s^+ \eta$ channels were considered, and the vector exchanged were the light vectors $\rho,\omega, K^{*}$. 
Here we also consider the exchange of heavy vectors and add $D_s^+\pi^0$ to the coupled channels. 
While $D_s^+\pi^0$ does not couple to the $DK$ isospin $I=0, \frac{1}{\sqrt{2}} (D^+K^0 + D^0 K^+)$, it couples to the individual components, and through the difference of mass between the charged and neutral kaons it can lead to the isospin violating mode of the $D_{s0}^*(2317)$. 
This source of isospin violation was early identified in Ref.~\cite{Achasov:1979xc} and plays a role in the present process as shown in Refs.~\cite{Guo:2008gp,Liu:2012zya,Guo:2018kno,Achasov:2025hkv}.
\begin{figure}[b]
	\begin{center}
		\includegraphics[width=0.6\linewidth]{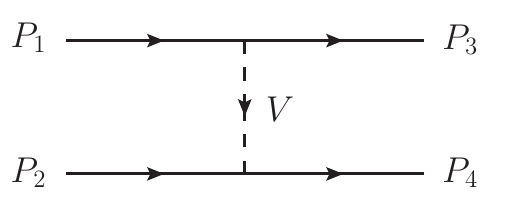}
		\vspace{-0.25cm}
		\caption{Diagrammatic representation of vector exchange as a source of the interaction between two pseudoscalars.}
		\label{fig:diagram1}
	\end{center}
\end{figure}

Our picture for the interaction is depicted in Fig.~\ref{fig:diagram1}.
It is driven by vector exchanges between two pseudoscalars, and all that is needed is the vertex $VPP$, which is given by the Lagrangian
\begin{equation}\label{eq:L}
	\mathcal{L}_{VPP}=-ig\,\langle [P,\partial_\mu P]\,V^\mu\rangle,
\end{equation}
where $g=\frac{M_V}{2 f_\pi}$ with $M_V\approx800 \mev$ and $f_\pi=93 \mev$.
The pseudoscalar $(P)$ and vector $(V)$ are the quark $q_{i} \bar{q}_{j}$ matrices written in terms of the physical mesons as Eq.~\eqref{eq:Pmatrix},
\begin{equation}\label{eq:Pmatrix}
    P =\scalebox{0.7}{
        $
        \left(
        \begin{array}{cccc}
            \frac{1}{\sqrt{2}}\pi^0 + \frac{1}{\sqrt{3}}\eta + \frac{1}{\sqrt{6}}\eta^{\prime} & \pi^+ & K^+  &  \bar D^0\\[2mm]
            \pi^- & -\frac{1}{\sqrt{2}}\pi^0 + \frac{1}{\sqrt{3}}\eta + \frac{1}{\sqrt{6}}\eta^{\prime} & K^0  & D^-\\[2mm]
            K^- & \bar{K}^0 & ~-\frac{1}{\sqrt{3}}\eta + \sqrt{\frac{2}{3}}\eta^{\prime}~  & D_s^-\\[2mm]
            D^0 & D^+ & D_s^+  & \eta_c\\
        \end{array}
        \right)$
    },
\end{equation}
where the standard $\eta-\eta^{\prime}$ mixing of Ref.~\cite{Bramon:1992kr} has been used, and 
\begin{equation}\label{eq:Vmatrix}
V =
\left(
\begin{array}{cccc}
	\frac{1}{\sqrt{2}}\rho^0 + \frac{1}{\sqrt{2}}\omega  & \rho^+ & K^{*+}  &  \bar D^{*0}\\[2mm]
	\rho^- & -\frac{1}{\sqrt{2}}\rho^0 + \frac{1}{\sqrt{2}}\omega  & ~K^{*0}~  & D^{*-}\\[2mm]
	K^{*-} & \bar{K}^{*0} & \phi  & D_s^{*-}\\[2mm]
	D^{*0} & D^{*+} & D_s^{*+}  & J/\psi \\
\end{array}
\right).
\end{equation}
By labelling the channels $D^0 K^+ (1)$, $D^+ K^0 (2)$, $D_s^+ \eta (3)$, $D_s^+ \pi^0 (4)$, we obtain the transition potential between these channels corresponding to the mechanism of Fig.~\ref{fig:diagram1} as
\begin{align}\label{eq:Vij}
    V_{11} &= C_{11}f(p_{i}), & V_{12} &= V_{21} = C_{12}f(p_{i}), \nonumber\\[2mm]
	V_{22} &= C_{22}f(p_{i}), &  V_{13} &= V_{31} = C_{13}f(p_{i}) + C_{13}^{\prime}g(p_{i}),\nonumber\\[2mm]
	V_{23} &= V_{32} = V_{13}, & V_{14} &= V_{41} = C_{14}f(p_{i}) + C_{14}^{\prime}g(p_{i}),\nonumber\\[2mm]
	V_{24} &= V_{42} = -V_{14}, & V_{34} &= V_{43} = 0, \nonumber\\[2mm]
	V_{33} &= 0, & V_{44} &= 0, 
\end{align}
where
\begin{equation}\label{eq:figi}
	\begin{aligned}
		f(p_{i}) &=g^{2}\;(p_1+p_3)\cdot(p_2+p_4),\\
		g(p_{i}) &=g^{2}\;(p_1+p_4)\cdot(p_2+p_3),
	\end{aligned}
\end{equation}
and the coefficients $C_{ij}, C^{\prime}_{ij}$ are given by
\begin{align}\label{eq:Cij}
    C_{11} &= -\frac{1}{2}\left( \frac{1}{m_\rho^2} + \frac{1}{m_\omega^2} \right), & C_{12} &=- \frac{1}{m_\rho^2}, \nonumber\\[2mm]
	C_{13} &= \frac{2}{\sqrt{3}}\frac{1}{m^2_{K^*}}, &
	C_{22} &= -\frac{1}{2}\left( \frac{1}{m_\rho^2} + \frac{1}{m_\omega^2} \right), \nonumber\\[2mm] 
	C_{23} &= \frac{2}{\sqrt{3}} \frac{1}{m^2_{K^*}}, & C_{14} &= \frac{1}{\sqrt{2}}\frac{1}{m^2_{K^*}},\nonumber\\[2mm]
	C_{13}^{\prime} &= \frac{1}{\sqrt{3}}\frac{1}{m^2_{D^*}},& C_{14}^{\prime} &= \frac{1}{\sqrt{2}}\frac{1}{m^2_{D^*}}.	 
\end{align}
Projected over $S$-wave, the factors  $f(p_{i}), g(p_{i})$ of Eq.~\eqref{eq:figi} give
\begin{align}\label{eq:figi2}
		(p_1+p_3) \cdot (p_2+p_4)&\to \frac{1}{2}\big[3s-(m_{1}^2+m_{2}^2+m_{3}^2+m_{4}^2) \nonumber\\
		&-\frac{1}{s}(m_{1}^2-m_{2}^2) (m_{3}^2-m_{4}^2)\big], \\
		(p_1+p_4) \cdot (p_2+p_3)&\to \frac{1}{2}\big[3s-(m_{1}^2+m_{2}^2+m_{3}^2+m_{4}^2) \nonumber\\
		&+\frac{1}{s}(m_{1}^2-m_{2}^2) (m_{3}^2-m_{4}^2)\big],
\end{align}
where $m_1, m_2$ correspond to the  $D$($D_s$) and light initial mesons and $m_3, m_4$ to the $D$($D_s$) and light final mesons respectively.

Once the potential has been constructed, the scattering amplitude $T$ matrix is given by
\begin{equation}\label{eq:TBS}
	T = [1-VG]^{-1} V\, ,
\end{equation}
where $G$ is the diagonal loop function $G = \text{diag} (G_1, G_2, G_3, G_4)$, with $G_i$ regularized in the cutoff method as
\begin{eqnarray}\label{eq:Gcut}
	G_i (s) = \!\!\int_{|{\vec q}\,| <q_{\rm max}} \dfrac{\mathrm{d}^3q}{(2\pi)^3} \, \dfrac{\omega_1 + \omega_2}{2 \omega_1 \omega_2} \dfrac{1}{s-(\omega_1 + \omega_2)^2+i\epsilon},~~~~~
\end{eqnarray}
with $\omega_i = \sqrt{{\vec{q}}^{\,2}+m_{i}^2  }$, for the two particles involved in the meson loop function. 
The magnitude of $q_{\rm max}$ reflects the range of the interaction in momentum space~\cite{Gamermann:2009uq,Song:2022yvz} and is fine tuned to get the pole of the $T$ matrix at the mass of the $D_{s0}^*(2317)$. 
Since all channels are closed, except the $D_s^+ \pi^0$ which is open for decay, the pole must be searched in the second Riemann sheet where for the channel $D_s^+\pi^0$ we change
\begin{eqnarray}\label{eq:G2cut}
	G \to G^{II}=G + i \dfrac{1}{4\pi \sqrt{s}}\;q; ~q=\dfrac{\lambda^{1/2}(s,m_{D_{s}^{+}}^2,m_{\pi^{0}}^{2})}{2\sqrt{s}}.~~~~~
\end{eqnarray}
Then we obtain the couplings of the resonance to each channel using
\begin{eqnarray}\label{eq:Tij}
   \lim_{s\to M_R^2} T_{ij} \simeq \dfrac{g_i\, g_j}{s-M_{R}^{2}+iM_{R}\,\Gamma},
\end{eqnarray}
which provides one coupling with an arbitrary sign, say $g^2_{1}$, and the rest with relative good sign through
\begin{equation}
	g_{j}=g_{1} \lim_{s \to M_R^2} \, (s-M_R^2)\; \dfrac{T_{1j}}{T_{11}}.
\end{equation} 
Once this is done, we have four methods to obtain the width of the $D_{s0}^*(2317)$. 
First, the pole is obtained at $\sqrt{s}+i\Gamma/2$ in the complex plane.
This already gives us the width of the state.

A second method to determine the width is to plot $|T_{ij}|^2 $ and look at the width of the distribution at half the strength of the peak.

A third method is to use the standard formula
\begin{eqnarray}\label{eq:Gamma}
	\Gamma=\dfrac{1}{8\pi}\; \dfrac{1}{M_{D_{s0}^{*}}^{2}}\,\left | g_{D_s^+\pi^0} \right |^2 \,q_{\pi^{0}},
\end{eqnarray}
with $q_{\pi^{0}}$ the momentum of the pion in the $D_{s0}^*(2317)^+ \to D_s^+ \pi^0$ decay, and $g_{D_s^+ \pi^0}$ the coupling of the resonance to $D_s^+ \pi^0$ obtained via Eq.~\eqref{eq:Tij}.
	
A fourth method is described below using a triangle diagram.

\subsection{Triangle mechanism to determine the $D_{s0}^*(2317)$ decay width}

\begin{figure}[tb]
	\begin{center}
		\includegraphics[width=0.95\linewidth]{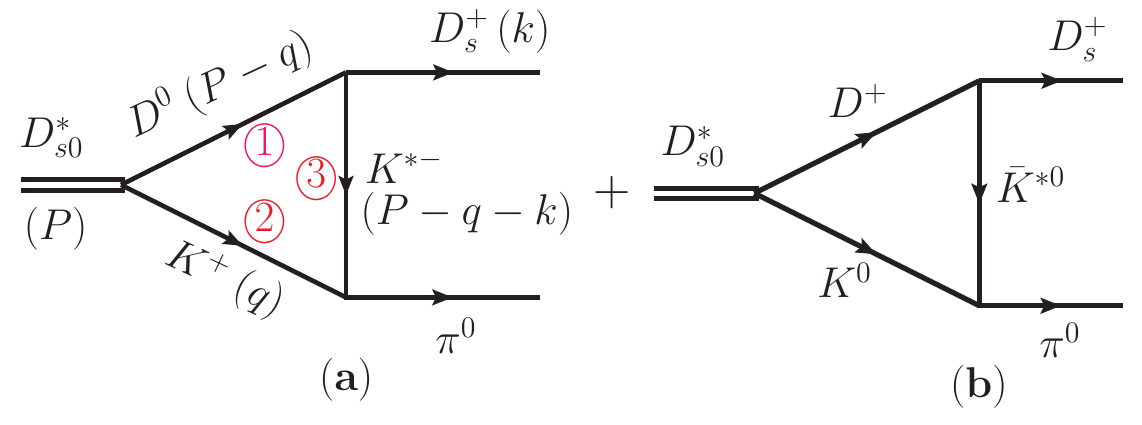}\\
		\includegraphics[width=0.95\linewidth]{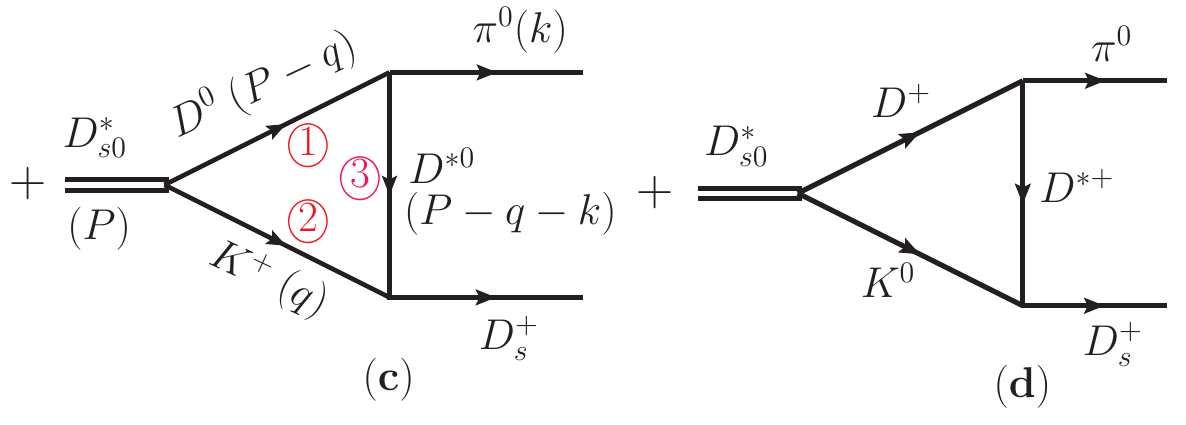}
		\vspace{-0.3cm}
		\caption{Diagrams entering the evaluation of the coupling of $D_{s0}(2317)^+$ to $D_s^+ \pi^0$, with the momenta of the particles in parenthesis.}
		\label{fig.diagram2}
	\end{center} 
\end{figure}

This method is often used to determine decay widths into channels of lesser importance to be considered as a coupled channel ~\cite{Lu:2021irg}. 
Considering the important components $D^0 K^+, D^+ K^0, D_s^+ \eta$, the diagrams to be considered are shown in Fig.~\ref{fig.diagram2}. 
The diagrams involing $\eta D_s^+$ vanish due to $C$-parity or flavor conservation.
The structure of these diagrams is the same and we have
\begin{equation}\label{eq:it}
	\begin{aligned}[b]
		-i t^{(s)} = -i g_{D_{s0}^{*}}^{(s)} &\int \dfrac{\mathrm{d}^4 q}{(2\pi)^4} \;\dfrac{i}{q^2 - m_2^2 + i\epsilon} \\
		&\cdot \dfrac{i}{(P-q)^2 - m_1^2 + i\epsilon} \,(-i)\,V^{(s)},
	\end{aligned}
\end{equation}
with $s = a, b, c, d$ being the label of the diagrams in Fig.~\ref{fig.diagram2}, and $V^{(s)}$ corresponding to the $D^{0}K^{+} (D^+ K^0) \to D_{s}^{+}\pi^{0}$ transition potential evaluated above in Eq.~\eqref{eq:Vij}, where, as a novelty, we do not take the off shell propagator $[(\vec{P}-\vec{q}-\vec{k}\,)^2-m_{3}^{2}]^2 \to \frac{1}{-m_{3}^{2}}$ but keep the full propagator. 
On the other hand, the $D^0$ and $K^+$ propagators, for the particles stemming from the $D_{s0}^*$ decay, are close to on shell and then we take the positive energy part of the propagator, the first term in the decomposition
\begin{equation}\label{eq:propagator}
	\dfrac{1}{q^2 - m^2 + i\epsilon} = \dfrac{1}{2\,\omega(\vec{q}\,)} \big[\dfrac{1}{q^0 - \omega(\vec{q}\,) + i\epsilon} - \dfrac{1}{q^0 + \omega(\vec{q}\,) - i\epsilon}\big].
\end{equation}
With this it is easy to perform the $q^0$ integration of Eq.~\eqref{eq:it} using the residues in Cauchy's integration, and we obtain finally
\begin{align}\label{eq:ts}
		t^{(s)} = & g_{D_{s0}^{*}}^{(s)} \int \dfrac{\mathrm{d}^3 q}{(2\pi)^3} \,
		\dfrac{1}{2\,\omega_1(\vec{q}\,)} 
		\dfrac{1}{2\,\omega_2(\vec{q}\,)} 
		\dfrac{1}{2\,\omega_3(\vec{q}+\vec{k}\,)} \nonumber\\
		& \times \dfrac{- M_{V}^{2}}{P^0 - \omega_1(\vec{q}\,) - \omega_2(\vec{q}\,) + i\epsilon}
		\; V^{(s)} \; \Theta (q_{\rm{max}} - |\vec{q}\,|) \nonumber\\
		& \times \left\{ \dfrac{1}{P^0 - k^0 - \omega_2(\vec{q}\,) - \omega_3(\vec{q}+\vec{k}\,) + i\epsilon} \right.\nonumber\\
		&\left. \quad\quad + \dfrac{1}{k^0 - \omega_1(\vec{q}\,) - \omega_3(\vec{q}+\vec{k}\,) + i\epsilon} \right\}.
\end{align}
We set $-M_{V}^{2}\,V^{(s)} =\tilde{V}^{(s)}$, which is given by
\begin{equation}
	\begin{aligned}\label{eq:tildeV}
\tilde{V}^{(a)} &= -\frac{1}{\sqrt{2}} \;f(p_i), &\quad&  \tilde{V}^{(b)} &= \frac{1}{\sqrt{2}} \; f(p_i),\\
\tilde{V}^{(c)} &= -\frac{1}{\sqrt{2}} \;g(p_i), &\quad& \tilde{V}^{(d)} &= \frac{1}{\sqrt{2}} \;g(p_i),
	\end{aligned}
\end{equation}
with $f(p_i)$ and $g(p_i)$ given in Eq.~\eqref{eq:figi}. 
Note that in Eq.~\eqref{eq:ts} we have introduced $\Theta (q_{\rm{max}} - |\vec{q}\,|)$. 
This is demanded in our approach since the $T$ matrix for $PP \to PP$ obtained in our approach, with the cutoff regularization of Eq.~\eqref{eq:Gcut} renders a $T$ matrix of the type $T(\vec{q}, \vec{q}^{\,\prime})=T\, \Theta (q_{\rm{max}} - |\vec{q}\,|)\cdot \Theta (q_{\rm{max}} - |\vec{q}^{\,\prime}|)$~\cite{Gamermann:2009uq}. 

The $D_{s0}^{*} \to D_s^+\pi^0$ transition amplitude is now given by
\begin{eqnarray}\label{eq:t5}
	t=t^{(a)}+t^{(b)}+t^{(c)}+t^{(d)}.
\end{eqnarray}
Then using Eq.~\eqref{eq:Gamma} with $| g_{D_s^+\pi^0}|^2 \to | t |^2$, we obtain the decay width of $D^*_{s0}(2317)^+$ to $D_s^+ \pi^0$.

We can see that $\tilde{V}^{(a)}$ and $\tilde{V}^{(b)}$ have opposite signs, and if the particles in the same isospin multiplet were the same, diagrams (a) and (b) would cancel.
The same happens with diagrams (c) and (d).
In this case, isospin would be conserved. The consideration of physical masses leads to the isospin breaking.

\section{Radiative decay of $D_{s0}^*(2317) \to D_s^* \gamma$}
\subsection{Charged current}
\begin{figure}[b]
\begin{center}
	\includegraphics[width=1\linewidth]{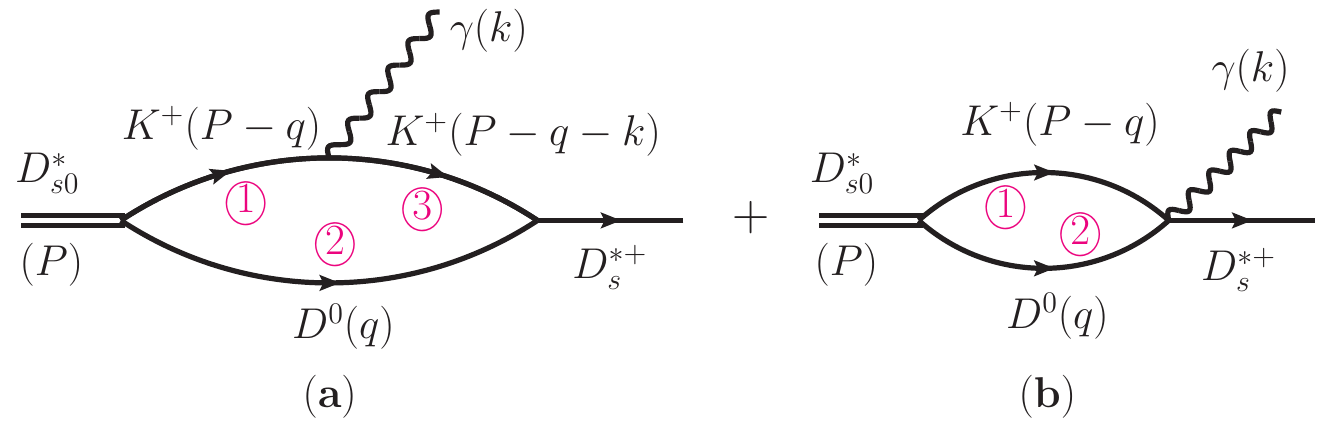}\\
	\includegraphics[width=1\linewidth]{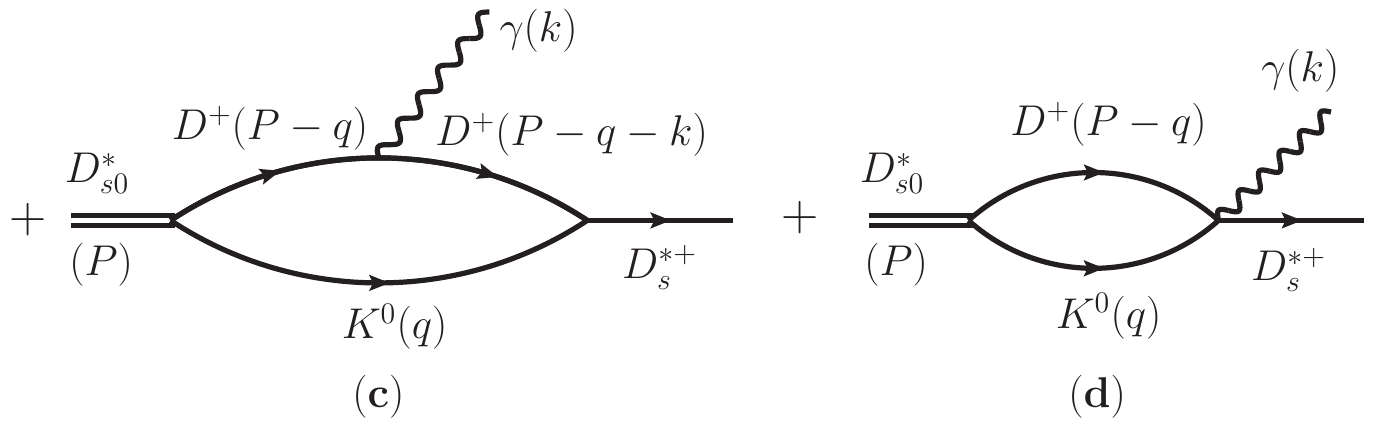}\\
	\includegraphics[width=1\linewidth]{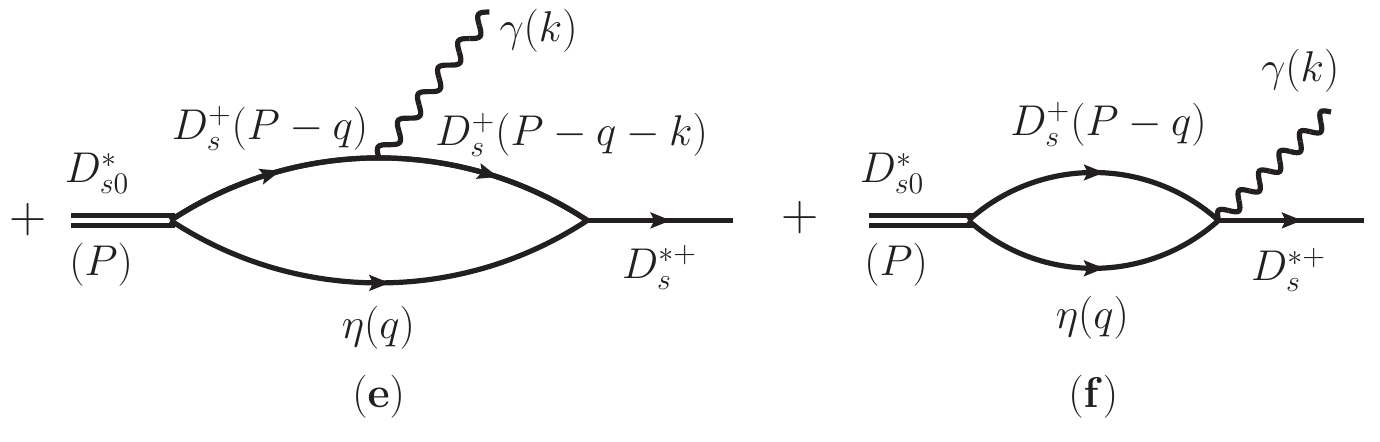}
	\caption{Diagrams involved in the $D_{s0}^{*}(2317)^+$ decay to $D_s^{*+} \gamma$, involving the charge coupling of the photons to the mesons. In parenthesis the momenta of the particles.}
	\label{fig.diagram3}
\end{center}
\end{figure}

Similarly to what has been done in the former section, we also use now the triangle diagrams to evaluate this width. 
The diagrams needed are shown in Fig.~\ref{fig.diagram3}.

\begin{figure}[tb]
	\begin{center}
		\includegraphics[width=0.55\linewidth]{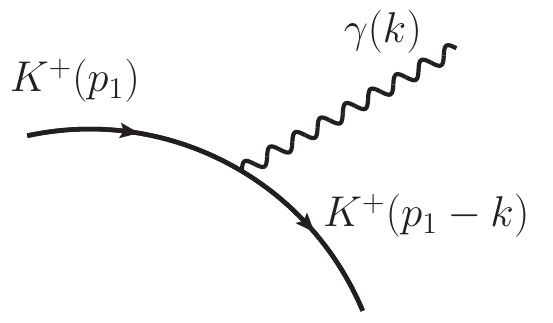}
		\vspace{-0.2cm}
		\caption{Diagrams for the charge coupling of a photon to a meson.}
		\label{fig.diagram4}
	\end{center}
\end{figure}

\begin{figure}[tb]
	\begin{center}
		\includegraphics[width=0.96\linewidth]{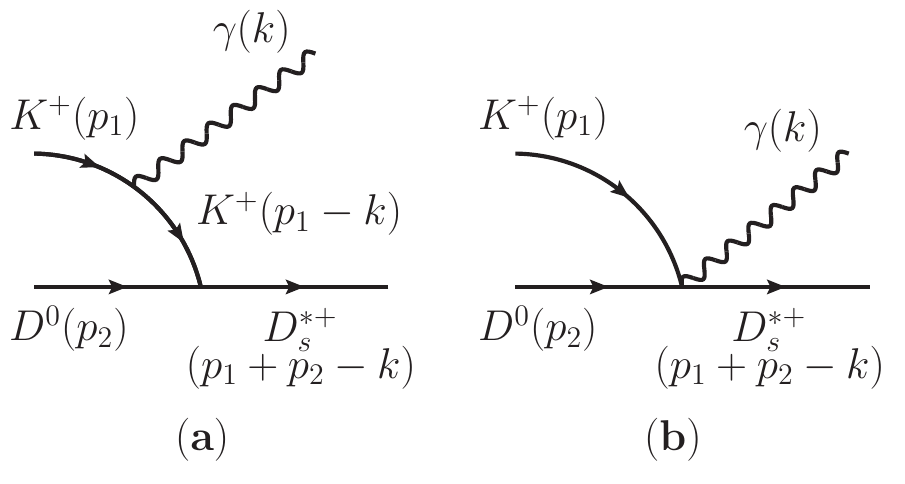}
		\vspace{-0.3cm}
		\caption{Meson in flight $(a)$ and contact term $(b)$ required by gauge invariance.}
		\label{fig.diagram5}
	\end{center}
\end{figure}

The photon coupling to the mesons involving the meson charge corresponding to the diagram of Fig.~\ref{fig.diagram4} is given by
\begin{equation}
	\mathcal{L} = i e Q \;(K^+ \partial_ \mu K^- - K^- \partial_\mu K^+) \;\epsilon^\mu(\gamma),
\end{equation}
where $K^+, K^-$ stand for the $K^+, K^-$ fields, $e>0$ is the electron charge ($e^{2}/4\pi=1/137$) and $Q$ is the charge of the charged meson, $Q=1$ in Fig.~\ref{fig.diagram4}. 
Once we reach this point, we have to face the issue of gauge invariance. In Ref.~\cite{Gamermann:2007bm} a method was used employing Feynman parametrization to account for gauge invariance. 
However, here we have to change the strategy, since in Eq.~\eqref{eq:ts} we must include the factor $\Theta (q_{\rm{max}} - |\vec{q}\,|)$. 
Then we introduce a contact term, Fig.~\ref{fig.diagram5}(b), to satisfy gauge invariance when we have the diagram of Fig.~\ref{fig.diagram5}(a).
In the diagram of Fig.~\ref{fig.diagram5}(a), one needs the lower vertex of the type $PPV$, which is given by the Lagrangian of Eq.~\eqref{eq:L}. Then we obtain
\begin{equation}
	\begin{aligned}[b]
		-i t'^{(a)} =& -i e Q g \;(2\,\vec{p}_1 - \vec{k})_\mu \,\epsilon^\mu(\gamma) \\
		&\cdot \dfrac{1}{(\vec{p}_1 - \vec{k}\,)^2 - m_K^2 + i\epsilon} \;
		\epsilon_\nu(D_s^*)\;(\vec p_1 - \vec p_2 - \vec{k}\,)^\nu.
	\end{aligned}
\end{equation}
This term requires a contact term to implement gauge invariance.
\begin{equation}
	-i t'^{(b)} = i e Q g \;\epsilon_{\mu}(\gamma) \,\epsilon^{\mu}(D_{s}^{*}).
\end{equation}
We shall work in the Coulomb gauge for the photons, $\epsilon^0(\gamma)=0, \, \vec{\epsilon}\cdot\vec{k}=0$, so we only have transverse photons. 
We also neglect the $\epsilon^{0}$ component of the $D_{s}^{*}$ since the three momentum is small compared to its mass (see appendix of Ref.~\cite{Sakai:2017hpg}), and then we have the contact terms corresponding to the diagrams (b), (d), (f) of Fig.~\ref{fig.diagram3},
\begin{align}\label{eq:-it}
	-i t^{(\rm con)}(K^{+} D^{0} \to \gamma D_{s}^{*+}) &= -i e Q g \;\vec{\epsilon}\,(\gamma) \cdot \vec{\epsilon}\,(D_{s}^{*}), \nonumber\\
	-i t^{(\rm con)}(D^+ K^{0} \to \gamma D_{s}^{*+}) &= i e Q g \; \vec{\epsilon}\,(\gamma) \cdot \vec{\epsilon}\,(D_{s}^{*}), \\
	-i t^{(\rm con)}(D_{s}^{+} \eta \to \gamma D_{s}^{*+}) &= -i e \frac{1}{\sqrt{3}} \,e Q g \;\vec{\epsilon}\,(\gamma) \cdot \vec{\epsilon}\,(D_{s}^{*}). \nonumber
\end{align}

Next we proceed as done in the triangle of the former section, taking the positive energy part of the propagators of the two mesons stemming from the $D_{s0}^{*}$ and the full propagator of the meson merging to the $D_{s}^{*}$ state in the diagrams. 
Then we use Cauchy's theorem to perform the $q^0$ integration and obtain for diagram (a) of Fig.~\ref{fig.diagram3},
\begin{equation}
	\begin{aligned}
	\tilde{t}^{\,(a)} = & g_{D_{s0}^{*},K^{+}D^{0}}\;e Q g \int \dfrac{\mathrm{d}^3 q}{(2\pi)^3}\; 
	\dfrac{1}{2\,\omega_1(\vec{q}\,)} \;
	\dfrac{1}{2\,\omega_2(\vec{q}\,)} 
	\dfrac{1}{2\,\omega_3(\vec{q}+\vec{k}\,)} \\
	& \times \dfrac{1}{P^0 - \omega_1(\vec{q}\,) - \omega_2(\vec{q}\,) + i\epsilon}
	 \cdot \Theta (q_{\rm{max}} - |\vec{q}\,|) \\
	& \times \left\{ \frac{1}{P^0 - k^0 - \omega_2(\vec{q}\,) - \omega_3(\vec{q}+\vec{k}\,) + i\epsilon} \right.\\
	&\left. \quad\quad + \dfrac{1}{k^0 - \omega_1(\vec{q}\,) - \omega_3(\vec{q}+\vec{k}\,) + i\epsilon} \right\}\\
	&\times(2\,q + k)_{i}\; (2\, q+k)_{j} \;\epsilon_{i}(\gamma)\;
\epsilon_{j}(D_{s}^{*}) \; \Theta (q_{\rm max}- |\vec q \,|),
	\end{aligned}
\end{equation}
where $1, 2, 3$ correspond to $K^{+}, D^{0}, K^{+}$.

Since $\int \mathrm{d}^{3}q \,q_{i} \;f(\vec{q},\vec{k}\,)$ is  proportional to $k_{i}$ and $\epsilon_{i}(\gamma)\,k_{i}=0$, in the factor $(2q+k)_{i}\,(2q+k)_{j}$ only the $4 \,q_i \,q_j$ term will contribute, and we have
\begin{equation}\label{eq:A}
	\int \dfrac{\mathrm{d}^3 q}{(2\pi)^3} \, q_i \, q_j \;f(\vec{q}, \vec{k}\,) = A \delta_{ij} + B k_i k_j,
\end{equation}
where the term $B$ will not contribute and the term $A$ is given by
\begin{equation}
	A = \frac{1}{2\, {\vec{k}}^2} \int \frac{\mathrm{d}^3q}{(2\pi)^3} \left[ \vec{q}^{\,2} \vec{k}^{\,2} - (\vec{q} \cdot \vec{k}\,)^2 \right] f(\vec{q},\vec{k}\,).
\end{equation}
Hence, taking this into account we reach the final formula
\begin{align}
		\tilde{t}^{\,(a)} &=  g_{D_{s0}^{*},K^{+}D^{0}}\; e Q g \;\dfrac{2}{{\vec{k}}^2} \, \epsilon_{i}(\gamma)\,
		\epsilon_{j}(D_{s}^{*}) \nonumber\\
		&\times \int \frac{\mathrm{d}^3 q}{(2\pi)^3} \left[ {\vec{q}}^{\,2}\, {\vec{k}}^{2} - (\vec{q} \cdot \vec{k}\,)^2 \right] 
		\frac{1}{2\,\omega_1(\vec{q}\,)} 
		\frac{1}{2\,\omega_2(\vec{q}\,)} 
		\frac{1}{2\,\omega_3(\vec{q}+\vec{k}\,)}  \nonumber\\
		& \times \frac{1}{P^0 - \omega_1(\vec{q}\,) - \omega_2(\vec{q}\,) + i\epsilon}
		\cdot \Theta (q_{\rm{max}} - |\vec{q}\,|)  \nonumber\\
		& \times \left\{ \frac{1}{P^0 - k^0 - \omega_2(\vec{q}\,) - \omega_3(\vec{q}+\vec{k}\,) + i\epsilon} \right.  \nonumber\\
		&\left. \quad\quad + \frac{1}{k^0 - \omega_1(\vec{q}\,) - \omega_3(\vec{q}+\vec{k}\,) + i\epsilon} \right\}.
\end{align}

The sum over polarizations of $\gamma$ and $D_s^*$ is easily done in the Coulomb gauge by taking into account that for transverse photons
\begin{equation}
	\sum_{\mathrm{pol}} \epsilon_i(\gamma) \; \epsilon_j(\gamma) = \delta_{ij} - \dfrac{k_i \,k_j}{{\vec{k}}^2},
\end{equation}
while for the $D_s^*$ we simply have $\sum_{\mathrm{pol}} \epsilon_i(D_{s}^{*}) \; \epsilon_j(D_{s}^{*}) = \delta_{ij}$. Hence, calling
\begin{equation}
	\tilde{t}^{\,(a)} = \tilde{t}^{\prime(a)} \;\epsilon_{i}(\gamma) \;\epsilon_{i}(D_{s}^{*}), 
\end{equation}
\begin{equation}
	\sum_{\text{pol}} | \tilde{t}^{(a)} |^2 = 2 | \tilde{t}^{\prime(a)} |^2.
\end{equation}

\subsection{Contact terms}
Next we evaluate the contact terms of Fig.~\ref{fig.diagram3}(b),(d),(f).
Referring to diagram \ref{fig.diagram3}(b) we have
\begin{align}
		-i\tilde{t}^{\,(b)} = & -ig_{D_{s0}^{*},K^{+}D^{0}} \; e Q g \nonumber\\[1mm]
		&\times \int \frac{\mathrm{d}^4 q}{(2\pi)^4} \;\frac{i}{(P-q)^2-m_{1}^2+i\epsilon}\; \frac{i}{q^2-m_{2}^2+i\epsilon} \nonumber\\[1mm]
		&\times (-i)\,\vec{\epsilon}\,(\gamma) \cdot \vec{\epsilon}\,(D_{s}^{*})\;
		\Theta (q_{\rm{max}} - |\vec{q}\,|),
\end{align}
which is finally evaluated with the result
\begin{align}
		\tilde{t}^{\,(b)} = & g_{D_{s0}^{*},K^{+}D^{0}}\; e Q g \;\epsilon_{i}(\gamma)\,\epsilon_{i}(D_{s}^{*}) \int \frac{\mathrm{d}^3 q}{(2\pi)^3} \frac{\omega_{1}(\vec{q}\,)+\omega_{2}(\vec{q}\,)}{2\,\omega_{1}(\vec{q}\,) \;\omega_{2}(\vec{q}\,)} \nonumber\\[1mm]
		&\times \frac{1}{M_{D_{s0}^{*}}^2-(\omega_{1}(\vec{q}\,)+\omega_{\vec{2}}(q)\,)^2+i\epsilon} \; \Theta (q_{\rm{max}} - |\vec{q}\,|).
\end{align}
The other diagrams of Fig.~\ref{fig.diagram3} are formally evaluated following the rules:
\begin{enumerate}
	\item[ i)] Diagram (c) is like (a) with the prescription
	\begin{enumerate}
		\item[1)] it has a relative minus sign;
		\item[2)] the particles $1, 2, 3$ are now $D^{+}, K^{0}, D^{+}$;
		\item[3)] the $D_{s0}^{*}$ coupling is now $g_{D_{s0}^{*}, \,D^+K^0}$.
	\end{enumerate}
	\item[ ii)] Diagram (e) is like (a) with
	\begin{enumerate}
		\item[1)] the particles $1, 2, 3$ are now $D_s^{+}, \eta, D_s^{+}$;
		\item[2)] there is the same sign but an extra coefficient $\frac{1}{\sqrt{3}}$;
		\item[3)] the $D_{s0}^{*}$ coupling  is now  $g_{D_{s0}^{*},\, D_s^+ \eta}$.
	\end{enumerate}
	\item[iii)] Diagram (d) is like (b) with
	\begin{enumerate}
		\item[1)] it has a relative minus sign;
		\item[2)] the particles $1, 2$ are now $D^+, K^0$;
		\item[3)] the $D_{s0}^{*}$ coupling is $g_{D_{s0}^{*}, \,D^+K^0}$.
	\end{enumerate}
	\item[iv)] Diagram (f) is like (b) with
	\begin{enumerate}
		\item[1)] has an extra coefficient $\frac{1}{\sqrt{3}}$ and the same sign;
		\item[2)] the particles $1, 2$ are now $D_s^+, \eta$;
		\item[3)] the $D_{s0}^{*}$ coupling is $g_{D_{s0}^{*}, \eta D_{s}^+}$.
	\end{enumerate}
\end{enumerate}
The sum of all the amplitudes is given by
\begin{align}
		\tilde{t} &= \tilde{t}^{\,(a)} + \tilde{t}^{\,(b)} + \tilde{t}^{\,(c)} + \tilde{t}^{\,(d)} + \tilde{t}^{\,(e)} + \tilde{t}^{\,(f)} \nonumber\\
		&= \epsilon_i(\gamma) \;\epsilon_i(D_s^*) \left\{\tilde{t}^{\prime(a)} + \tilde{t}^{\prime(b)} + \tilde{t}^{\prime(c)} + \tilde{t}^{\prime(d)} \right. \nonumber\\
		&\left.\quad +\tilde{t}^{\prime(e)} +\tilde{t}^{\prime(f)} 
		\right\},
\end{align}
and the width corresponding to all these diagrams is
\begin{equation}\label{eq:Gamma1}
	\Gamma_{D_{s0}^{*} \to \gamma D_s^{*+}} = \frac{1}{8\pi} \; \frac{1}{m_{D_{s0}^{*}}^2} \; \sum_{\text{pol}} | \tilde{t}|^2 \,q_{\gamma},
\end{equation}
with 
\begin{equation}\label{eq:sumt}
	 \sum_{\text{pol}} | \tilde{t} |^2 =2\, \big|\sum_{s}  \tilde{t}^{\prime(s)} \big|^2,
\end{equation}
\begin{equation}\label{eq:qgamma}
	q_{\gamma}=\dfrac{\lambda^{1/2}(M_{D_{s0}^{*}}^{2}, 0, M_{D_{s}^{*+}}^{2})}{2 \,M_{D_{s0}^{*}}}.
\end{equation}

\subsection{Anomalous terms} 
\begin{figure*}[tb]
	\begin{center}
		\includegraphics[width=0.96\linewidth]{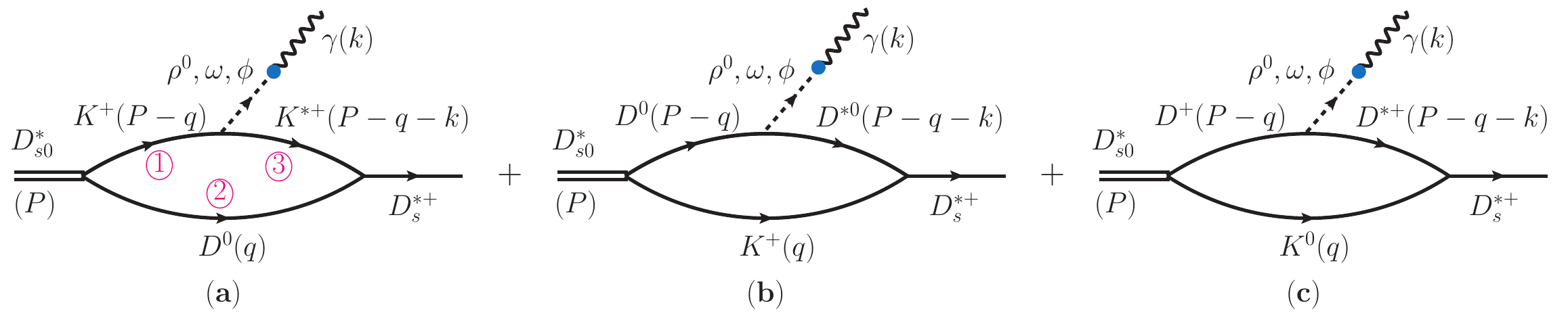}\\
		\includegraphics[width=0.96\linewidth]{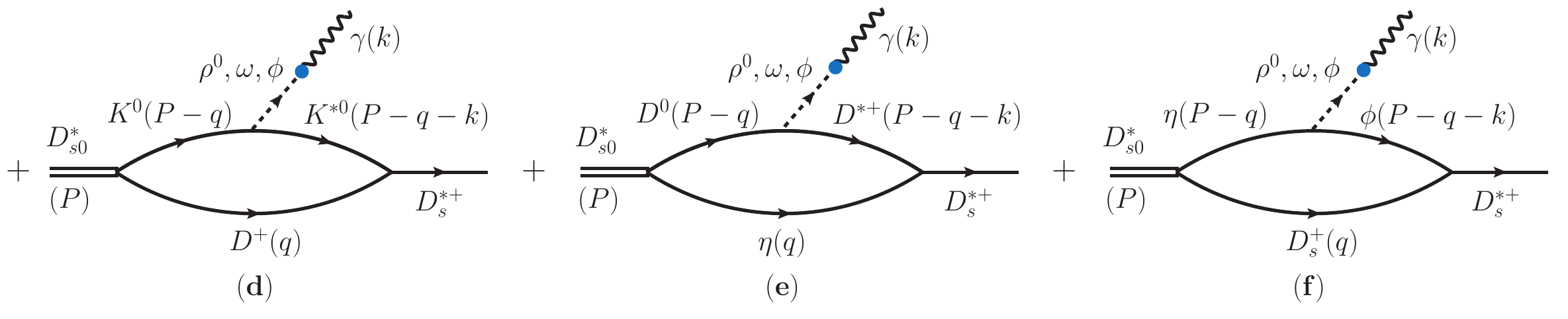}
		\vspace{-0.1cm}
		\caption{Diagrams involving anomalous couplings involved in the $D_{s0}^{*}(2317)^+ \to D_{s}^{*+}\gamma$ radiative decay.}
		\label{fig.diagram6}
	\end{center}
\end{figure*}

We consider now the terms in which the off shell pseudoscalar meson merging into $D_s^{*+}$ is substituted by a vector meson. 
This gives rise to the diagrams of Fig.~\ref{fig.diagram6}.
We need now an extra Lagrangian coupling two vectors to a pseudoscalar meson, 
\begin{equation}
	\mathcal{L}_{VVP} = \frac{G^{\prime}}{\sqrt{2}} \; \varepsilon^{\mu\nu\alpha\beta} \,\langle \partial_\mu V_\nu \partial_\alpha V_\beta P \rangle,
\end{equation}
with $V$ the vector matrix of the Eq.~\eqref{eq:Vmatrix} and $G^{\prime}$ given by~\cite{Bramon:1992kr,Oset:2002sh,Ramos:2013mda}
\begin{equation}
	G^{\prime}=\dfrac{3 \, M_{V}^2}{16\, \pi^2 f^3}\simeq 14 ~\mathrm{GeV}^{-1}.
\end{equation}
To connect the $VVP$ Lagrangians with a final photon, we use vector meson dominance (VMD), producing $\rho^0, \omega, \phi$ which later convert into a photon via the conversion Lagrangian~\cite{Bando:1984ej,Bando:1987br,Meissner:1987ge,Nagahiro:2008cv}
\begin{equation}
	\begin{aligned}[b]\label{eq:LVgamma}
	\mathcal{L}_{V\gamma}&=-4\,f^2 eg \; A_{\mu} \langle \widetilde{Q} \,V^{\mu}  \rangle\\
	&=-4\,M_{V}^2 \;\dfrac{e}{g}\; A_{\mu}\langle \widetilde{Q}\, V^{\mu}  \rangle,
	\end{aligned}
\end{equation}
where $V$ is again the matrix of Eq.~\eqref{eq:Vmatrix} reduced to a $3\times3$ of the light sector, $A_\mu$ the photon field and $e<0$ (we will change the sign later to adjust to the first part with the charge currents). $\widetilde{Q}$ in Eq.~\eqref{eq:LVgamma} is the matrix for the charge of the quarks
\begin{equation}
\widetilde{Q} = \frac{1}{3}
\left( \begin{array}{ccc}
	2 & 0 & 0 \\
	0 & -1 & 0 \\
	0 & 0 & -1
\end{array} \right).
\end{equation}
The conversion Lagrangian gives rise to a vertex
\begin{equation}\label{eq:Lvertex}
	\begin{aligned}[b]
		-it&=-i\frac{M_{V}^{2}\,e}{g}\;A_{\mu}\; V^{\mu}\;C_{\gamma},
	\end{aligned}
\end{equation}
where
\begin{equation}
	\begin{aligned}[b]
C_{\gamma}=\begin{Bmatrix}
	\frac{1}{\sqrt{2}},~~\rho^0 
	\\
	\frac{1}{3}\frac{1}{\sqrt{2}},~\omega 
	\\
	-\frac{1}{3},~~\phi
	
\end{Bmatrix}.
	\end{aligned}
\end{equation}
We should note that now all terms involving the anomalous coupling and VMD satisfy gauge invariance.
If we look at the diagram of Fig.~\ref{fig.diagram6}(a) we find for the $K^{+}VK^{*+}$ ($V= \rho^0, \omega, \phi$) vertex
\begin{equation}
	\mathcal{L} = \frac{G^{\prime}}{\sqrt{2}} \; \varepsilon^{\mu\nu\alpha\beta} \;\partial _{\mu}K^{*-}_{\nu}\left\{ \partial _{\alpha} \left(\frac{\rho^0}{\sqrt{2}}+\frac{\omega}{\sqrt{2}}\right)_{\beta}+\partial _{\alpha}\phi_{\beta}\right\}K^{+},
\end{equation}
\begin{equation}\label{eq:vertices}
	\to \frac{G'}{\sqrt{2}}\; \varepsilon^{\mu\nu\alpha\beta} \;i(P-q)_\mu \;\epsilon_\nu(K^*) \;ik_\alpha 
	\left\{ \begin{matrix}
		\frac{1}{\sqrt{2}}, & \rho^0   \\
		\frac{1}{\sqrt{2}}, & \omega   \\
		1, & \phi
	        \end{matrix} \right\}
	\epsilon_{\beta}(V),
\end{equation}
where we have removed a term $\varepsilon^{\mu\nu\alpha\beta}\,k_{\nu}\,k_{\alpha}$. 
The three momenta are relatively small compared to the masses of the vector mesons and we take $\epsilon^{0}(V)=0$, thus having only the spatial $\vec{\epsilon}\,(V)$ polarization vectors (see the accuracy of this approximation in Ref.~\cite{Dias:2025izv}). 
Then the zero index in $\varepsilon^{\mu\nu\alpha\beta}$ can only be $\nu$ or $\alpha$ and then using $\varepsilon^{0ijk}=\varepsilon^{ijk}$ we obtain
\begin{equation}
	\begin{aligned}
\frac{G'}{\sqrt{2}}\varepsilon^{ijk} &\left\{(P^0-q^0)\,k_{j}-k^{0}\,(P-q)_{j}\right\}\\
&\times \epsilon_i(K^*) \;\epsilon_{k}(V)
\left\{ \begin{matrix}
	\frac{1}{\sqrt{2}}, & \rho^0   \\
	\frac{1}{\sqrt{2}}, & \omega   \\
	1, & \phi
\end{matrix} \right\},
	\end{aligned}
 \end{equation}
where all indices are meant to be contravariant.
Similarly, for the lower vertex of Fig.~\ref{fig.diagram6}(a) we obtain
\begin{equation}
\mathcal{L} \to \varepsilon^{\mu\nu\alpha\beta}\;\partial_{\mu}\,{K^{*+}}_{\nu}\, \partial_{\alpha}\,{D_{s}^{*-}}_\beta \,D^0,
\end{equation}
and now the zero index of $\varepsilon^{\mu\nu\alpha\beta}$ corresponds to the external $D_{s}^{*}$, and we get
\begin{equation}\label{eq:LL}
	\mathcal{L}=-\frac{G'}{\sqrt{2}}\; \varepsilon^{i^{\prime}j^{\prime}k^{\prime}} \; (q+k)_{j^{\prime}} \,(P-k^0) \;\epsilon_i^{\prime}(K^*)\; \epsilon_{k^{\prime}}(D_{s}^{*}).
\end{equation}
Combining Eqs.~\eqref{eq:vertices} and~\eqref{eq:LL} and summing $\sum_{\text{pol}}\epsilon_{i}(K^*)\,\epsilon_{i^{\prime}}(K^*)=\delta_{ii^{\prime}}$ and using
\begin{equation}
	\sum_{i}\varepsilon^{ijk}\, \varepsilon^{ij^{\prime}k^{\prime}}=\delta_{jj^{\prime}}\,\delta_{kk^{\prime}}-\delta_{jk^{\prime}}\,\delta_{kj^{\prime}},
\end{equation}
we find the combined term
\begin{align}\label{eq:sum}
	\frac{G^{\prime 2}}{2}&\frac{P^0-k^0}{(P-q-k)^2-m_{K^{*}}^2}\left\{[(P^0-q^0)k_j-k^0(P-q)_j]\right. \nonumber\\[1mm]
	&\cdot (q+k)_j \; \epsilon_k(V) \;\epsilon_k(D_{s}^*)-[(P^0-q^0)k_j-k^0(P-q)_j] \nonumber\\[1mm]
	&\cdot \left.\epsilon_{j}(D_{s}^*)\;\epsilon_{k}(V)\,(q+k)_k\right\}\cdot\left\{ \begin{matrix}
		\frac{1}{\sqrt{2}} ,&\rho^0   \\[1mm]
		\frac{1}{\sqrt{2}},& \omega   \\[1mm]
		1,&\phi
	\end{matrix} \right\}.
\end{align}
The first term in Eq.~\eqref{eq:sum} already has the structure $\epsilon_k(V)\epsilon_k(D_{s}^{*})$. 
In the second term $\epsilon_k(V)k_k$ will vanish after $V\rightarrow \gamma$ conversion, the terms linear in $\vec{q}$ become proportional to $\vec{k}$ after the integration and vanish, and only the $q_j q_k$ term remains, which via Eq.~\eqref{eq:A} gives rise to a term $A\delta_{jk}$ which provides a contribution.

Then, following the same steps as done before to evaluate the loop integrals using Cauchy integration to evaluate $q^0$ integration and considering the $V-\gamma$ conversion vertices of Eq.~\eqref{eq:Lvertex}, where we already change the sign of $e$, to make it positive, we finally obtain
\begin{align}\label{eq:tildeta}
		\tilde{t}_{\rm An}^{\,(a)}=&\frac{e}{g}\frac{1}{3}\frac{G^{\prime 2}}{2}\,g_{D_{s0},K^+D^0}\;\epsilon_j(\gamma)\;\epsilon_j(D_s^*) \nonumber\\[1mm]
		\times\int&\frac{\mathrm{d}^3q}{(2\pi)^3}\;\frac{1}{2\,\omega_1(\vec{q}\,)}\; 
		\frac{1}{2\,\omega_2(\vec{q}\,)} 
		\frac{1}{2\,\omega_3(\vec{q}+\vec{k}\,)}  \nonumber\\[1mm]  
		&\times\frac{P^0-k^0}{P^0-\omega_1(\vec{q}\,)-\omega_2(\vec{q}\,)+i\epsilon}\;\Theta (q_{\rm{max}} - |\vec{q}\,|)  \nonumber\\[1mm]
		&\times\left\{ \frac{D_1}{P^0 - k^0 - \omega_2(\vec{q}\,) - \omega_3(\vec{q}+\vec{k}\,) + i\epsilon} \right.   \nonumber\\[1mm]
		&\left. \quad\quad + \frac{D_2}{k^0 - \omega_1(\vec{q}\,) - \omega_3(\vec{q}+\vec{k}\,) + i\epsilon} \right\},
	\end{align}
where $D_{1}, D_{2}$ are given by
\begin{align}\label{eq:D1D2}
		D_1=&[P^0-\omega_2(\vec{q}\,)]\; \vec{k} \cdot (\vec{q}+\vec{k})+k^0\, \vec{q} \cdot (\vec{q}+\vec{k}\,)\nonumber\\
		&-\frac{k^0}{2\,{\vec{k}}^2}\;[{\vec{q}}^{\,2}\, {\vec{k}}^{\,2}-(\vec{q}\cdot\vec{k}\,)^2],\\
		D_2=&\omega_1(\vec{q}\,)\,\vec{k} \cdot (\vec{q}+\vec{k}\,)+k^0\, \vec{q} \cdot (\vec{q}+\vec{k}\,) \nonumber\\
		&-\frac{k^0}{2\,{\vec{k}}^2} \; [{\vec{q}}^{\,2} \, {\vec{k}}^{\,2}-(\vec{q}\cdot\vec{k}\,)^2].
\end{align}
We should note the difference between $D_1$ and $D_2$ is that $q^0$ in Eq.~\eqref{eq:sum} has to be substituted by $\omega_2(\vec{q}\,)$ for the first term of Eq.~\eqref{eq:propagator} and by $P^0-\omega_1(\vec{q}\,)$ for the second term in the Cauchy integration using residues.

The rest of the diagrams of Fig.~\ref{fig.diagram6} are automatically evaluated with the simple changes which are shown in Table~\ref{table1}. 
\begin{table}[t]
	\centering
	\caption{Coefficients and particles entering the different diagrams of Fig.~\ref{fig.diagram6}.}\label{table1} 
	\begin{tabular}{|c|c|c|c|}
		\hline
		Diagram     &Factor & particles $1,~2,~3$ & $D_{s0}^*$ coupling \\[1mm]
		\hline
		(a) &$\frac{1}{3}$  & $K^+, D^0, K^{*+}$ & $g_{D_{s0}^*, \,D^0 K^+}$ \\[2mm]
		(b) &$\frac{2}{3}$  & $D^0, K^+, D^{*0}$ & $g_{D_{s0}^*, \,D^0 K^+}$ \\[2mm]
		(c) &$-\frac{1}{3}$ & $D^+, K^0, D^{*+}$ & $g_{D_{s0}^*, \,D^+ K^0}$ \\[2mm]
		(d) &$-\frac{2}{3}$ & $K^0, D^+, K^{*0}$ & $g_{D_{s0}^*, \,D^+ K^0}$ \\[2mm]
		(e) &$\frac{1}{3\sqrt{3}}$ & $D_s^+,\eta, D_s^{*+}$ & $g_{D_{s0}^*, \,D_s^+\eta}$ \\[2.5mm]
		(f) &$\frac{1}{3\sqrt{3}}$  & $\eta, D_s^+, \phi$ & $g_{D_{s0}^*,  \,D_s^+\eta}$ \\[2mm]
		\hline
	\end{tabular}
\end{table}

The anomalous terms are now
\begin{equation}
	\begin{aligned}[b]
		\tilde{t}_{\rm An}=&\tilde{t}_{\rm An}^{\,(a)}+\tilde{t}_{\rm An}^{\,(b)}+\tilde{t}_{\rm An}^{\,(c)}+\tilde{t}_{\rm An}^{\,(d)}+\tilde{t}_{\rm An}^{\,(e)}+\tilde{t}_{\rm An}^{\,(f)}\\
		=&\epsilon_{i}(\gamma)\;\epsilon_{i}(D_{s}^{*+})\;\sum_{s}\tilde{t}_{\rm An}^{\,\prime (s)},
	\end{aligned}
\end{equation}
which defines $\tilde{t}_{\rm An}^{\,\prime (s)}$.  
The width due to the mechanism is obtained via Eqs.~\eqref{eq:Gamma1}-\eqref{eq:qgamma} substituting $\tilde{t}^{\,(s)}$ by $\tilde{t}_{\rm An}^{\,(s)}$. 
Then summing the $\tilde{t}^{\,(s)}$ and $\tilde{t}_{\rm An}^{\,(s)}$ amplitudes and using these equations, we obtain the total width.

\section{Results}
\subsection{Strong decay of $D_{s0}^{*}(2317)$ to $D_s^+\pi^0$}
We summarize the results of the coupled channels part in Table~\ref{tab:pole_couplings}.
We fine tuned $q_{\mathrm{max}}$ to obtain the mass of the $D_{s0}^{*}(2317)$ at the physical mass and obtained $q_{\mathrm{max}}=687~\mathrm{MeV}$, in line with the result of Ref.~\cite{Ikeno:2023ojl}.
In Table~\ref{tab:pole_couplings} we also see the couplings of $D_{s0}^{*}$ to the different channels. 
The couplings to $D^{0}K^{+}, D^{+}K^{0}, D_{s}^{+}\eta$ are in line with those of Ref.~\cite{Ikeno:2023ojl} and the coupling to the isospin violating $D_{s}^+\pi^0$ channel is new, and, as expected, small compared to the other couplings.
\begin{table*}[htbp]
	\centering
	\caption{Pole position and couplings with $q_{\text{max}} = 687$ MeV. [in units of MeV]}
	\label{tab:pole_couplings}
	\renewcommand{\arraystretch}{1.2}
	\begin{tabular}{|c|c|c|c|c|}
		\hline
		$\sqrt{s_{p}}$ & $g_{D^{0}K^{+}}$ & $g_{D^{+}K^{0}}$ & $g_{D_{s}^{+}\eta}$ & $g_{D_{s}^{+}\pi^{0}}$ \\[2mm]
		$(2317.74 + 0.03537i)$ & $8252.26 - 69.15i$ & $8129.49 + 75.70i$ & $-6312.66 - 2.50i$ & $-174.59 + 39.95i$ \\
		\hline
	\end{tabular}
\end{table*}

\begin{figure}[tb]
	\begin{center}
		\includegraphics[width=0.9\linewidth]{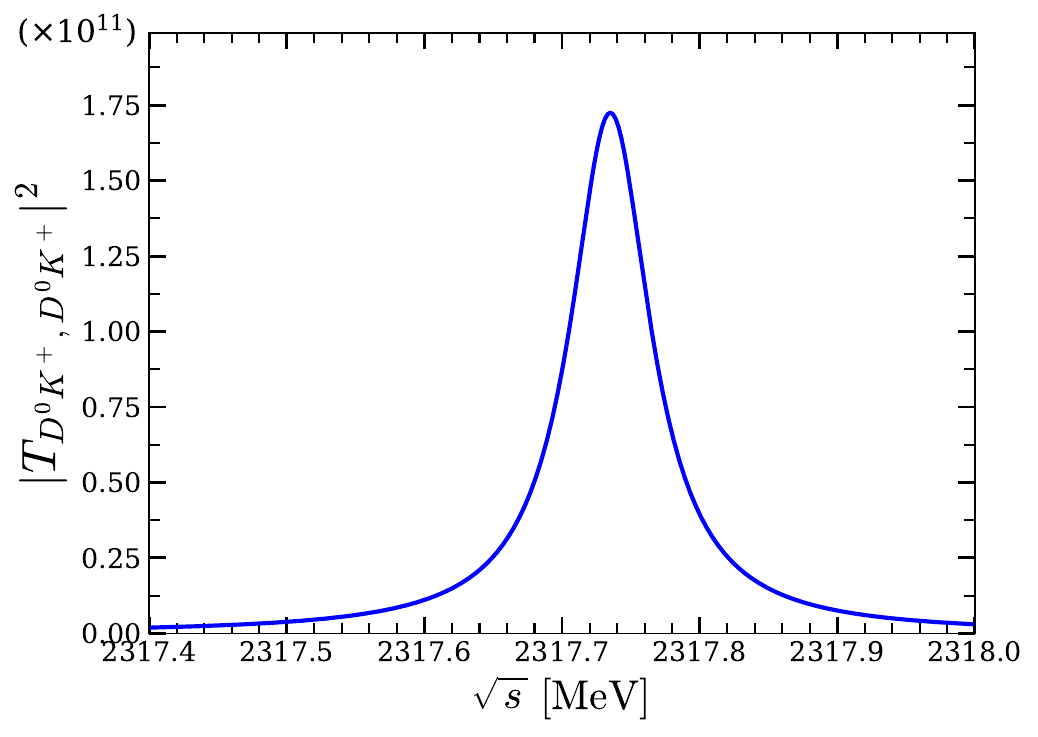}
		\vspace{-0.2cm}
		\caption{$|T_{11}|^2 $ as a function of $\sqrt{s}$.}
		\label{fig.diagram7}
	\end{center} 
\end{figure}

From two times the imaginary part of the pole position in Table~\ref{tab:pole_couplings}, we find 
\begin{equation}
	\begin{aligned}[b]
		\Gamma^{(\text{pole})}_{D_{s0}^{*}(2317)}=70.75 \kev.
	\end{aligned}
\end{equation}
Another method to evaluate the width is using the $g_{D_s^+\pi^0}$ coupling of Table~\ref{tab:pole_couplings} and Eq.~\eqref{eq:Gamma}. 
Then we find
\begin{equation}
	\begin{aligned}[b]
		\Gamma^{(\text{coupling})}_{D_{s0}^{*}(2317)}=70.77\kev.
	\end{aligned}
\end{equation}
The third method is using $|T_{ij} |^2$ to see the width of the peak. 
We show $|T_{11}|^2 $ in Fig.~\ref{fig.diagram7}. 
From the width at half the strength of the peak we find
\begin{equation}
	\begin{aligned}[b]
		\Gamma^{(T_{11})}_{D_{s0}^{*}(2317)}=70.77 \kev.
	\end{aligned}
\end{equation} 

Finally, using the fourth method with the triangle diagrams, we find
\begin{equation}
	\begin{aligned}[b]
		\Gamma^{(\text{triangle})}_{D_{s0}^{*}(2317)}=70.83 \kev.
	\end{aligned}
\end{equation}
As we can see, the four methods agree remarkably. 

We should recall that according to Eq.~\eqref{eq:tildeV} there should be a cancellation of diagrams $(a)$ and $(b)$ and $(c)$ and $(d)$ of Fig.~\ref{fig.diagram2} if the $K^+, K^0 \,(K^{*+}, K^{*0})$ masses were equal.
However, this is not the case when the masses are different, and the cancellation is only partial, giving rise to some isospin violation. 
Note, however, that there is another source, since by solving the coupled channels equations with different $K^+, K^0$ masses, the coupling of the $D_{s0}^{*}$ to $D^0 K^+$ and $D^+ K^0$, are not equal, as seen in Table~\ref{tab:pole_couplings}. 
The cancellation of terms indicated above only occurs if these couplings are equal. 
Hence, there are two sources of isospin violation in our approach.

\subsection{Radiative decays}
If we consider the charge current terms of Fig.~\ref{fig.diagram3} ignoring the contact terms, we obtain
\begin{equation}
	\begin{aligned}[b]
		\Gamma^{(\text{current})}_{D_{s0}^{*}\rightarrow D_{s}^{*+}\gamma}=0.55 \kev.
	\end{aligned}
\end{equation}
If in addition we consider the contact terms, we obtain
\begin{equation}
	\begin{aligned}\label{eq:Gammacc}
		\Gamma^{(\text{current+contact})}_{D_{s0}^{*}\rightarrow D_{s}^{*+}\gamma}=1.36 \kev.
	\end{aligned}
\end{equation}
The result of Eq.~\eqref{eq:Gammacc} is bigger than the one found in Ref.~\cite{Gamermann:2007bm} of $0.475^{+0.831}_{-0.290} \kev$, but nearly compatible within errors. 
In the evaluation of Ref.~\cite{Gamermann:2007bm} the couplings to the $D^0 K^+$ and $D^+ K^0$ were considered equal, and furthermore the couplings obtained there were a bit smaller $g_{D_{s0}^{*}, \,D^0 K^+}=7358 \mev$. 
Considering the ratio of couplings, the rate of Ref.~\cite{Gamermann:2007bm} could reach the value of $0.6 \kev$, still small compared to the number of Eq.~\eqref{eq:Gammacc}, indicating that the second source of isospin violation from having different couplings to the $D^0 K^+$ and $D^+ K^0$ components is very important. 

The anomalous terms are individually not small, in particular from diagram (b) of Fig.~\ref{fig.diagram6} we obtain $ \Gamma^{(\text{anomalous, b})}=0.5 \kev$. 
Other terms provide individual widths of $0.1 \kev$, and diagrams (e), (f) involving $\eta$ particles are basically negligible at the level of $4\times 10^{-3} \kev$. 
The curious thing is that there is a large cancellation between all terms of Fig.~\ref{fig.diagram6}, and the final contribution of all the anomalous terms is
\begin{equation}
	\begin{aligned}
		\Gamma^{(\text{anomalous})}_{D_{s0}^{*}\rightarrow D_{s}^{*+}\gamma}=5.5\times 10^{-4} \kev,
	\end{aligned}
\end{equation}
such that when we sum all the terms from the current and anomalous couplings, we obtain a final width
\begin{equation}
	\begin{aligned}
		\Gamma_{D_{s0}^{*}\rightarrow D_{s}^{*+}\gamma}=1.33 \kev.
	\end{aligned}
\end{equation}
Then the ratio of the radiative decay width to the $D_{s}^{+}\pi^{0}$ decay is
\begin{equation}
	\begin{aligned}
		\frac{\Gamma_{D_{s0}^{*}\rightarrow D_{s}^{*+}\gamma}}{\Gamma_{D_{s0}^{*}\rightarrow D_{s}^{+}\pi^0}}=1.88 \%. 
	\end{aligned}
\end{equation}
This result of about $2 \%$ is smaller than that found in Ref.~\cite{Belle-II:2025dzk} of about $7\%$. 
In the next section we make some reflection about this result.

\section {Discussion and conclusions}
We have evaluated the strong decay of $D_{s0}^*(2317)^+$ to $D_s^+ \pi^0$ considering coupled channels of $D^0 K^+$, $D^+ K^0$, $D_s^+ \eta$ and $D_s^+ \pi^0$ to build a molecular state for the $D_{s0}^*(2317)^+$. 
The explicit consideration of  $D^0 K^+, D^+ K^0$ instead of the isospin $I=0$ component is necessary because the isospin forbidden $D_s^+ \pi^0$ decay mode appears thanks to the mass difference between charged and neutral kaons. 
We also applied a different method, using a conventional triangle mechanism to produce the $D_s^+ \pi^0$ decay mode from the main building blocks of the molecule $D^0 K^+, D^+ K^0, D_s^+ \eta$, and the results are remarkably similar.  
We obtain a decay width of about $71 \kev$, to which one would have to add a $\pi^0-\eta$ mixing contribution which we discuss below.
    
We have also calculated the radiative decay to $D_s^{*+} \gamma$, using the same information obtained from the molecular picture, both using charged currents and anomalous terms.  

At this point it is convenient to discuss uncertainties in the results which we address in the subsection below.

\subsection{Uncertainties}
We evaluate now uncertainties from different sources.
First in the strong decay we have in Eq.~\eqref{eq:figi} the factor $g^2 \equiv M_V^2/4f^2$, and for $M_V$ we have taken $800\mev$, which is an average between the vector mesons masses, but we keep $m_\rho, m_\omega, m_{K^*}$ different in Eqs.~\eqref{eq:Cij}.
We take now in Eqs.~\eqref{eq:Cij} all the vector meson masses equal to the average $M_V$.
With this and the same $q_{\rm max} =687 \mev$ we obtain the results in Table \ref{tab:new}, which are similar to those obtained before.
We note that the mass of the $D^*_{s0}(2317)$ state has changed a bit to the value of $2314\mev$.
We then change a bit $q_{\rm max}$ to regain the right mass, which is attained with $q_{\rm max}=675\mev$.
We recalculate the couplings with this new input and the results are also shown in Table \ref{tab:new}.
\begin{table*}[htbp]
	\centering
	\caption{Same as Table \ref{tab:pole_couplings} but with $m_i=M_V$ in Eq.~\eqref{eq:Cij}. [in units of MeV]}
	\label{tab:new}
	\renewcommand{\arraystretch}{1.2}
	\begin{tabular}{|c|c|c|c|c|}
		\hline
		$\sqrt{s_{p}} \;(q_{\rm max}=687)$ & $g_{D^{0}K^{+}}$ & $g_{D^{+}K^{0}}$ & $g_{D_{s}^{+}\eta}$ & $g_{D_{s}^{+}\pi^{0}}$ \\[2mm]
		$(2314.33 + 0.038i)$ & $8245.85 - 89.33i$ & $8128.56 + 97.34i$ & $-7447.25 - 3.55i$ & $-178.41 + 59.45i$ \\
		\hline
        $\sqrt{s_{p}}\;(q_{\rm max}=675)$ & $g_{D^{0}K^{+}}$ & $g_{D^{+}K^{0}}$ & $g_{D_{s}^{+}\eta}$ & $g_{D_{s}^{+}\pi^{0}}$ \\[2mm]
		$(2317.71 + 0.041i)$ & $8145.70 - 92.01i$ & $8032.44 + 100.62i$ & $-7384.42 - 3.83i$ & $-182.26 + 61.52i$ \\
		\hline
	\end{tabular}
\end{table*}
Of relevance to us is the new result for $\Gamma_{D_s\pi^0}$ which is now given by
\begin{equation}
	\begin{aligned}\label{eq:newResult}
\Gamma^{(\text{pole})}_{D_{s0}^{*}(2317)}&=81.58 \kev,\\[2mm]
\Gamma^{(\text{coupling})}_{D_{s0}^{*}(2317)}&=81.62\kev, \\[2mm]
\Gamma^{(\text{triangle})}_{D_{s0}^{*}(2317)}&=94.48\kev.
	\end{aligned}
\end{equation}
This gives us an idea of the uncertainties, of the order of $\pm 7\kev$, in an average with the former results of $\Gamma_{\rm strong}\simeq 77\kev$
\begin{equation}\label{eq:Gamstrong1}
   \Gamma_{\rm strong} = 77\pm 7 \kev.
\end{equation}

We come now to the radiative decay. Here we have the charged contribution proportional to $g=M_V/2f$.
If we take $M_V=770\mev$ (as for the $\rho$ meson) or $M_V=1020\mev$ (as for the $\phi$ meson), the decay widths that go as $g^2$ change from $1.3\kev$ to $1.2\kev$ ($M_V=770\mev$) or $2.2\kev$ ($M_V=1020\mev$).
The effect of the change of the $D^*_{s0}(2317)$ couplings is small.
Hence we can take
\begin{equation}
   \Gamma_{\rm rad.} = (1.7\pm 0.5) \kev.
\end{equation}

\subsection{Effect of $\pi^0-\eta$ mixing}
From the consideration of $u, d, s$ quark mass differences, there is a term for direct $\pi^0-\eta$ mixing \cite{Cho:1994zu} given by \cite{Cho:1994zu,Guo:2006fu}
\begin{equation}
   t_{\pi\eta} = \langle\pi^0|H|\eta\rangle = -0.003 \;{\rm GeV^2}.
\end{equation}
The effect of this mixing can be obtained \cite{Guo:2006fu} by multiplying the $D^*_{s0}(2317)$ coupling to $D_s^+ \eta$ by 
\begin{equation} 
   \dfrac{t_{\pi\eta}}{m^2_{\pi^0}-m^2_\eta} = 0.0107.
\end{equation}
This provides the coupling of $D^*_{s0}(2317)$ to $D_s^+ \pi^0$ from the mechanism of $\pi^0-\eta$ mixing.

By taking the coupling $g_{D^+_s \eta}$ of Table \ref{tab:pole_couplings}, we obtain an extra coupling to $D_s^+ \pi^0$
\begin{equation} 
   g'_{D_s^+ \pi^0}\simeq -67.55 \mev,
\end{equation}
which added to $g_{D^+_s \pi^0}$ of Table \ref{tab:pole_couplings} gives
\begin{equation} 
   g_{D^+_s \pi^0}+g'_{D_s^+ \pi^0}= -242.14+i39.95 \mev,
\end{equation}
and by using Eq.~\eqref{eq:Gamma}, we obtain the strong decay width of $D^*_{s0}(2317)$,
\begin{equation} \label{eq:Gamstrong2}
   \Gamma_{\rm strong}= 133 \kev,
\end{equation}
a large increase from the original value of $71 \kev$.

We can perform a different calculation by implement the $\pi^0-\eta$ mixing into the potential $V_{ij}$ of Eq.~\eqref{eq:Vij}.
One can take the non vanishing transitions $V_{13}, V_{23}$ (from $D^0 K^+, D^+ K^0$ to $D^+_s \eta$) and apply to them the $\pi^0-\eta$ transition correction.
This induces changes in $V_{14}, V_{24}$ which now become
\begin{equation}
	\begin{aligned}\label{eq:newResult}
V'_{14}&=V_{14} + V_{13} \;\dfrac{t_{\pi\eta}}{m^2_{\pi^0}-m^2_\eta},\\[2mm]
V'_{24}&=V_{24} + V_{23} \;\dfrac{t_{\pi\eta}}{m^2_{\pi^0}-m^2_\eta},
	\end{aligned}
\end{equation}
and then we recalculate the $T$ matrix and the new couplings.
We find now
\begin{equation} 
   g'_{D^+_s \pi^0}= -254.51+i57.43 \mev,
\end{equation}
by means of which
\begin{equation} \label{eq:Gamstrong3}
   \Gamma_{\rm strong}= 147 \kev.
\end{equation}
This number is very close ($8\%$ difference) to that of Eq.~\eqref{eq:Gamstrong2} and the difference can be taken as uncertainties.
We can take an average $140 \kev$ with uncertainty $\pm 7\kev$.
If we add this to the uncertainty from the coupled channels of Eq.~\eqref{eq:Gamstrong1} summed with quadrature we get
\begin{equation} \label{eq:Gamstrong4}
   \Gamma_{\rm strong}= (140 \pm 10) \kev.
\end{equation}

We have made a compilation of results for the strong and radiative decays of the $D_{s0}^*(2317)^+$ and show the results in Table \ref{tab:3}.
\begin{table*}[t]
	\centering
	\caption{Results for strong and radiative decays from different works. Methods are explained as follows. $c\bar{s}$: $q\bar{q}$ quark model; $c\bar{s}$ sum rules: sum rules assuming $\bar{q}q$ state;
	rel: relativistic; 
	pheno: phenomenological;
	pheno lattice: combined QCD lattice and phenomenology; 
	$^3P_0$: $^3P_0$ model; 
	CQM: constituent quark meson model;
	chQM: chiral quark model;
	HQET: heavy quark effective theory;
	HMchPT: heavy meson chiral perturbation theory; 
	$DK$ mol: $DK$ molecule;
	$DK +$ lattice: combined $DK$ and QCD lattice results; 
	$DK, D_s \eta$ mol: $DK, D_s \eta$ molecule;
	$DK, D_{s}\eta, D_{s}\pi^{0}$: molecule with these component.}
	\label{tab:3}
	\renewcommand{\arraystretch}{1.2}
	\begin{tabular}{ccccccccc|}
		\hline
		 method & \multicolumn{5}{|c|}{$c\bar{s}$} & pheno & \multicolumn{1}{c}{pheno lattice}& \\[1mm]
		 ref  & \multicolumn{1}{|c}{\cite{Godfrey:1985xj}} & \cite{Ebert:1997nk} & \cite{Godfrey:2003kg} & \cite{Ishida:2003gu} & \multicolumn{1}{c|}{\cite{Close:2005se}~~} & \cite{Azimov:2004xk}&\multicolumn{1}{c}{\cite{Yue:2025wcl}}&\\ [1mm]
		 strong\;[keV] &\multicolumn{1}{|c} {-} & ~$12$~ & ~~$10$~~ & ~$155\pm70$~ &\multicolumn{1}{c|}{-} & $109\pm 16$ & \multicolumn{1}{c}{63.0--209}&\\[1mm]
		 radiative\;[keV]~~~& \multicolumn{1}{|c}{~$0.125$~} & ~$0.19$~ & ~$1.9$~ & - & \multicolumn{1}{c|}{~~$1$~~}& 0.55 & \multicolumn{1}{c}{-}&\\[1mm]
		\hline 
         method &\multicolumn{2}{|c|}{$c\bar s$ sum rules}& $c\bar s$ heavy quark & rel $c\bar{s}$ &  $^3P_0$ & CQM & \multicolumn{1}{c}{chQM}&\\[1mm]
		 ref  & \multicolumn{1}{|c}{\cite{Wang:2006mf}} & \multicolumn{1}{c|}{\cite{Wei:2005ag}}& \cite{Colangelo:2003vg}& \cite{Goity:2000dk} & \cite{Lu:2006ry}& \cite{Liu:2006jx} &\multicolumn{1}{c}{\cite{Wang:2006fg}}&\\[1mm]
		 strong\;[keV] &\multicolumn{1}{|c} {-} &\multicolumn{1}{c|}{34--44}&6 &30.5&32&3.7--8.7&\multicolumn{1}{c}{1.5}&\\[1mm]
         radiative\;[keV]~~~& \multicolumn{1}{|c}{1.3--9.9} & \multicolumn{1}{c|}{-}&1 &-&-&0.54--1.4&\multicolumn{1}{c}{-}&\\[1mm]
		\hline 
        method &\multicolumn{2}{|c|}{HQET}&HQEF &HMchPT & & ~$DK$ mol~ & \multicolumn{1}{c}{~$DK +$ lattice~}&\\[1mm]
        ref  & \multicolumn{1}{|c}{\cite{Guo:2008gp}} & \multicolumn{1}{c|}{\cite{Fajfer:2015zma}} & \cite{Colangelo:2005hv} & \cite{Yang:2019cat} &&\cite{Faessler:2007gv}&\multicolumn{1}{c}{\cite{Liu:2012zya}}&\\[1mm]
		strong\;[keV] &\multicolumn{1}{|c} {~$180\pm 110~~$} &\multicolumn{1}{c|}{2.4--4.7~~}&6&5--12&&47 & \multicolumn{1}{c}{$133\pm 19$}&\\[1mm]
        radiative\;[keV]~~~& \multicolumn{1}{|c}{-} & \multicolumn{1}{c|}{-}&4--6&-&&0.47--1.41 &\multicolumn{1}{c}{-}&\\[1mm]
		\hline 
		method &\multicolumn{5}{|c|}{$DK, D_s \eta$ mol}& \multicolumn{2}{c}{$DK, D_{s}\eta, D_{s}\pi^{0}$}&\\[1mm]
		ref  & \multicolumn{1}{|c}{\cite{Gamermann:2007bm}} & \cite{Lutz:2007sk} &
		\cite{Cleven:2014oka} & \cite{Fu:2021wde} & \multicolumn{1}{c|}{\cite{Guo:2018kno}~~} & \cite{Achasov:2025hkv} & \multicolumn{1}{c}{This work}&\multicolumn{1}{c|}{with $\pi^0$-$\eta$ mix.}\\[1mm]
		strong\;[keV] &\multicolumn{1}{|c}{-}& 140 & $96\pm 19$ & $~132\pm 7~$ & \multicolumn{1}{c|}{~~104--116~~}& 95--130 & \multicolumn{1}{c}{$77\pm7$}& \multicolumn{1}{c|}{$140\pm10$}\\[1mm]
		radiative\;[keV]~~~&\multicolumn{1}{|c}{~$0.48^{+0.83}_{-0.29}$}&2.85&$9.4\pm 3.8$& $3.7\pm 0.3$&\multicolumn{1}{c|}{-}&-&\multicolumn{1}{c}{$1.7\pm0.5$}&-\\[1mm]
	\hline
\end{tabular} 
\end{table*}
To better interpret the results let us quote that Ref.~\cite{Cleven:2014oka} is updated in Ref.~\cite{Fu:2021wde}, Ref.~\cite{Guo:2008gp} in Ref.~\cite{Liu:2012zya} and Ref.~\cite{Lutz:2007sk} in Ref.~\cite{Guo:2018kno}.
We  can see that the molecular picture produces results for the strong decay of the order of $100 \kev$. 
On the other hand, models based on a $c \bar s$ quark component have larger uncertainties, but are roughly one order of magnitude smaller. 
When it comes to the radiative decay, the results based on $c \bar s$ quark model calculations show large variations, as much as one order of magnitude, and are generally smaller than the results coming from the molecular picture, but similar.  
Our calculation for the radiative decay included the contribution of anomalous terms, not considered in other approaches, but we could prove that their global contribution was negligible. 

If one looks at the ratio of the radiative to strong decay widths from the molecular picture, one finds in all cases ratios below $3 \%$, which would be in conflict with the experimental results of Ref.~\cite{Belle-II:2025dzk} of $7 \%$. 
Our calculation provides a ratio of $1.2\pm 0.4$, in line with other results and also smaller than the experimental ratio. 
One could invoke that from lattice QCD results one induces that the molecular picture has a strength  of about $70 \%$ of the $D_{s0}^*(2317)^+$ state \cite{MartinezTorres:2014kpc}. 
To account for that one could reduce the couplings of the resonance to the molecular components, but this would occur both in the strong and the radiative decays and the ratio would be the same. 
One can argue differently and say one could reduce the molecular contribution by $30 \%$ and add a $30 \%$ of the smaller contribution from a $q \bar q$ state. 
Since the radiative widths are similar in all cases, this would reduce the strong decay leaving the radiative decay the same and the ratio would increase by $25 \%$. Yet, this would not solve the problem in any case. 
On the other hand, the dispersion of results for the quark models, both in the radiative and strong decays, does not allow us to produce a meaningful ratio.
In a recent work \cite{Fu:2025lfo} the ratio is evaluated taking results from the molecular picture for the strong decay \cite{Liu:2012zya} and using experimental values of the ratio $R_2={\rm Br}(D_{s1}(2460)\to \gamma D_s)/{\rm Br}(D_{s1}(2460)\to \pi D^*_s)$.
The ratio varies from $0.01$ to $0.04$ depending on the values of $R_2$ from different experiments.
At this point it is also worth mentioning that the PDG \cite{ParticleDataGroup:2024cfk} still quotes a value for the ratio smaller than $0.059$ based on the experiment of Ref.~\cite{CLEO:2003ggt}, a results that is in conflict with the precise value from the Belle experiment \cite{Belle-II:2025dzk}. 
Although the numbers are similar, it will be useful to have more accurate result in the future. 
But, better than a ratio, what would help clarify the present theoretical debate would be to have precise measurements for the strong and the radiative decays independently. 
We are looking forward to this scenario in the future.  
Meanwhile, the present results, with the new methods of calculation, the combined study of the strong and radiative decays within the same framework, and the consistency with modern formulation of the molecular picture and the interaction of the their components, should be a valuable information to compare with these results when they are available. 

\vspace{0.4cm}
\section*{Acknowledgments}
We would like to thank Prof. Feng-Kun Guo for useful discussions.
This work is partly supported by the National Natural Science Foundation of China (NSFC) under Grants No. 12575081 and No. 12365019,
and by the Natural Science Foundation of Guangxi province under Grant No. 2023JJA110076,
and by the Central Government Guidance Funds for Local Scientific and Technological Development, China (No. Guike ZY22096024).
This work is partly supported the National Key R\&D Program of China (Grant No. 2024YFE0105200).
This work is also partly supported by the Spanish Ministerio de Economia y Competitividad (MINECO) and European FEDER funds under Contracts No. FIS2017-84038-C2-1-PB, PID2020-112777GB-I00, and by Generalitat Valenciana under contract PROMETEO/2020/023. 
This project has received funding from the European Union Horizon 2020 research and innovation program under the program H2020-INFRAIA-2018-1, grant agreement No. 824093 of the STRONG-2020 project.
Wen-Tao Lyu is supported by the Natural Science Foundation of Henan Grant No. 252300423951, the Zhengzhou University Young Student Basic Research Projects (PhD students) under Grant No. ZDBJ202522.

\bibliographystyle{a}
\bibliography{refs}

\begin{thebibliography}{65}%
\makeatletter
\providecommand \@ifxundefined [1]{%
 \@ifx{#1\undefined}
}%
\providecommand \@ifnum [1]{%
 \ifnum #1\expandafter \@firstoftwo
 \else \expandafter \@secondoftwo
 \fi
}%
\providecommand \@ifx [1]{%
 \ifx #1\expandafter \@firstoftwo
 \else \expandafter \@secondoftwo
 \fi
}%
\providecommand \natexlab [1]{#1}%
\providecommand \enquote  [1]{``#1''}%
\providecommand \bibnamefont  [1]{#1}%
\providecommand \bibfnamefont [1]{#1}%
\providecommand \citenamefont [1]{#1}%
\providecommand \href@noop [0]{\@secondoftwo}%
\providecommand \href [0]{\begingroup \@sanitize@url \@href}%
\providecommand \@href[1]{\@@startlink{#1}\@@href}%
\providecommand \@@href[1]{\endgroup#1\@@endlink}%
\providecommand \@sanitize@url [0]{\catcode `\\12\catcode `\$12\catcode `\&12\catcode `\#12\catcode `\^12\catcode `\_12\catcode `\%12\relax}%
\providecommand \@@startlink[1]{}%
\providecommand \@@endlink[0]{}%
\providecommand \url  [0]{\begingroup\@sanitize@url \@url }%
\providecommand \@url [1]{\endgroup\@href {#1}{\urlprefix }}%
\providecommand \urlprefix  [0]{URL }%
\providecommand \Eprint [0]{\href }%
\providecommand \doibase [0]{https://doi.org/}%
\providecommand \selectlanguage [0]{\@gobble}%
\providecommand \bibinfo  [0]{\@secondoftwo}%
\providecommand \bibfield  [0]{\@secondoftwo}%
\providecommand \translation [1]{[#1]}%
\providecommand \BibitemOpen [0]{}%
\providecommand \bibitemStop [0]{}%
\providecommand \bibitemNoStop [0]{.\EOS\space}%
\providecommand \EOS [0]{\spacefactor3000\relax}%
\providecommand \BibitemShut  [1]{\csname bibitem#1\endcsname}%
\let\auto@bib@innerbib\@empty
\bibitem [{\citenamefont {Wang}(2007{\natexlab{a}})}]{Wang:2006mf}%
  \BibitemOpen
  \bibfield  {author} {\bibinfo {author} {\bibfnamefont {Z.-G.}\ \bibnamefont {Wang}},\ }\bibinfo {title} {{Radiative decays of the $D_{s0}(2317)$, $D_{s1}(2460)$ and the related strong coupling constants}},\ \href {https://doi.org/10.1103/PhysRevD.75.034013} {\bibfield  {journal} {\bibinfo  {journal} {Phys. Rev. D}\ }\textbf {\bibinfo {volume} {75}},\ \bibinfo {pages} {034013} (\bibinfo {year} {2007}{\natexlab{a}})},\ \Eprint {https://arxiv.org/abs/hep-ph/0612225} {arXiv:hep-ph/0612225} \BibitemShut {NoStop}%
\bibitem [{\citenamefont {Kalashnikova}\ {\it et~al.}(2006)\citenamefont {Kalashnikova}, \citenamefont {Kudryavtsev}, \citenamefont {Nefediev}, \citenamefont {Haidenbauer},\ and\ \citenamefont {Hanhart}}]{Kalashnikova:2005zz}%
  \BibitemOpen
  \bibfield  {author} {\bibinfo {author} {\bibfnamefont {Y.}~\bibnamefont {Kalashnikova}}, \bibinfo {author} {\bibfnamefont {A.~E.}\ \bibnamefont {Kudryavtsev}}, \bibinfo {author} {\bibfnamefont {A.~V.}\ \bibnamefont {Nefediev}}, \bibinfo {author} {\bibfnamefont {J.}~\bibnamefont {Haidenbauer}},\ and\ \bibinfo {author} {\bibfnamefont {C.}~\bibnamefont {Hanhart}},\ }\bibinfo {title} {{Insights on scalar mesons from their radiative decays}},\ \href {https://doi.org/10.1103/PhysRevC.73.045203} {\bibfield  {journal} {\bibinfo  {journal} {Phys. Rev. C}\ }\textbf {\bibinfo {volume} {73}},\ \bibinfo {pages} {045203} (\bibinfo {year} {2006})},\ \Eprint {https://arxiv.org/abs/nucl-th/0512028} {arXiv:nucl-th/0512028} \BibitemShut {NoStop}%
\bibitem [{\citenamefont {Godfrey}(2003)}]{Godfrey:2003kg}%
  \BibitemOpen
  \bibfield  {author} {\bibinfo {author} {\bibfnamefont {S.}~\bibnamefont {Godfrey}},\ }\bibinfo {title} {{Testing the nature of the $D_{sJ}^*(2317)^+$ and $D_{sJ}(2463)^+$ States Using Radiative Transitions}},\ \href {https://doi.org/10.1016/j.physletb.2003.06.049} {\bibfield  {journal} {\bibinfo  {journal} {Phys. Lett. B}\ }\textbf {\bibinfo {volume} {568}},\ \bibinfo {pages} {254} (\bibinfo {year} {2003})},\ \Eprint {https://arxiv.org/abs/hep-ph/0305122} {arXiv:hep-ph/0305122} \BibitemShut {NoStop}%
\bibitem [{\citenamefont {Colangelo}\ and\ \citenamefont {De~Fazio}(2003)}]{Colangelo:2003vg}%
  \BibitemOpen
  \bibfield  {author} {\bibinfo {author} {\bibfnamefont {P.}~\bibnamefont {Colangelo}}\ and\ \bibinfo {author} {\bibfnamefont {F.}~\bibnamefont {De~Fazio}},\ }\bibinfo {title} {{Understanding $D_{sJ}(2317)$}},\ \href {https://doi.org/10.1016/j.physletb.2003.08.003} {\bibfield  {journal} {\bibinfo  {journal} {Phys. Lett. B}\ }\textbf {\bibinfo {volume} {570}},\ \bibinfo {pages} {180} (\bibinfo {year} {2003})},\ \Eprint {https://arxiv.org/abs/hep-ph/0305140} {arXiv:hep-ph/0305140} \BibitemShut {NoStop}%
\bibitem [{\citenamefont {Bardeen}\ {\it et~al.}(2003)\citenamefont {Bardeen}, \citenamefont {Eichten},\ and\ \citenamefont {Hill}}]{Bardeen:2003kt}%
  \BibitemOpen
  \bibfield  {author} {\bibinfo {author} {\bibfnamefont {W.~A.}\ \bibnamefont {Bardeen}}, \bibinfo {author} {\bibfnamefont {E.~J.}\ \bibnamefont {Eichten}},\ and\ \bibinfo {author} {\bibfnamefont {C.~T.}\ \bibnamefont {Hill}},\ }\bibinfo {title} {{Chiral Multiplets of Heavy - Light Mesons}},\ \href {https://doi.org/10.1103/PhysRevD.68.054024} {\bibfield  {journal} {\bibinfo  {journal} {Phys. Rev. D}\ }\textbf {\bibinfo {volume} {68}},\ \bibinfo {pages} {054024} (\bibinfo {year} {2003})},\ \Eprint {https://arxiv.org/abs/hep-ph/0305049} {arXiv:hep-ph/0305049} \BibitemShut {NoStop}%
\bibitem [{\citenamefont {Fayyazuddin}\ and\ \citenamefont {Riazuddin}(2004)}]{Fayyazuddin:2003aa}%
  \BibitemOpen
  \bibfield  {author} {\bibinfo {author} {\bibnamefont {Fayyazuddin}}\ and\ \bibinfo {author} {\bibnamefont {Riazuddin}},\ }\bibinfo {title} {{Some comments on narrow resonances $D_{s1}^{*} (2.46 ~\mathrm{GeV} / c^2)$ and $D_{s0} (2.317~\mathrm{GeV}/ c^2)$}},\ \href {https://doi.org/10.1103/PhysRevD.69.114008} {\bibfield  {journal} {\bibinfo  {journal} {Phys. Rev. D}\ }\textbf {\bibinfo {volume} {69}},\ \bibinfo {pages} {114008} (\bibinfo {year} {2004})},\ \Eprint {https://arxiv.org/abs/hep-ph/0309283} {arXiv:hep-ph/0309283} \BibitemShut {NoStop}%
\bibitem [{\citenamefont {Ishida}\ {\it et~al.}(2004)\citenamefont {Ishida}, \citenamefont {Ishida}, \citenamefont {Komada}, \citenamefont {Maeda}, \citenamefont {Oda}, \citenamefont {Yamada},\ and\ \citenamefont {Yamauchi}}]{Ishida:2003gu}%
  \BibitemOpen
  \bibfield  {author} {\bibinfo {author} {\bibfnamefont {S.}~\bibnamefont {Ishida}}, \bibinfo {author} {\bibfnamefont {M.}~\bibnamefont {Ishida}}, \bibinfo {author} {\bibfnamefont {T.}~\bibnamefont {Komada}}, \bibinfo {author} {\bibfnamefont {T.}~\bibnamefont {Maeda}}, \bibinfo {author} {\bibfnamefont {M.}~\bibnamefont {Oda}}, \bibinfo {author} {\bibfnamefont {K.}~\bibnamefont {Yamada}},\ and\ \bibinfo {author} {\bibfnamefont {I.}~\bibnamefont {Yamauchi}},\ }\bibinfo {title} {{The $D_{s}(2317)$ and $D_{s}(2463)$ mesons as scalar and axial vector chiralons in the covariant level classification scheme}},\ \href {https://doi.org/10.1063/1.1799786} {\bibfield  {journal} {\bibinfo  {journal} {AIP Conf. Proc.}\ }\textbf {\bibinfo {volume} {717}},\ \bibinfo {pages} {716} (\bibinfo {year} {2004})},\ \Eprint {https://arxiv.org/abs/hep-ph/0310061} {arXiv:hep-ph/0310061} \BibitemShut {NoStop}%
\bibitem [{\citenamefont {Azimov}\ and\ \citenamefont {Goeke}(2004)}]{Azimov:2004xk}%
  \BibitemOpen
  \bibfield  {author} {\bibinfo {author} {\bibfnamefont {Y.~I.}\ \bibnamefont {Azimov}}\ and\ \bibinfo {author} {\bibfnamefont {K.}~\bibnamefont {Goeke}},\ }\bibinfo {title} {{Decay properties of new $D$-mesons}},\ \href {https://doi.org/10.1140/epja/i2004-10010-4} {\bibfield  {journal} {\bibinfo  {journal} {Eur. Phys. J. A}\ }\textbf {\bibinfo {volume} {21}},\ \bibinfo {pages} {501} (\bibinfo {year} {2004})},\ \Eprint {https://arxiv.org/abs/hep-ph/0403082} {arXiv:hep-ph/0403082} \BibitemShut {NoStop}%
\bibitem [{\citenamefont {Colangelo}\ {\it et~al.}(2005)\citenamefont {Colangelo}, \citenamefont {De~Fazio},\ and\ \citenamefont {Ozpineci}}]{Colangelo:2005hv}%
  \BibitemOpen
  \bibfield  {author} {\bibinfo {author} {\bibfnamefont {P.}~\bibnamefont {Colangelo}}, \bibinfo {author} {\bibfnamefont {F.}~\bibnamefont {De~Fazio}},\ and\ \bibinfo {author} {\bibfnamefont {A.}~\bibnamefont {Ozpineci}},\ }\bibinfo {title} {{Radiative transitions of $D^{*}_{sJ}(2317)$ and $D_{sJ}(2460)$}},\ \href {https://doi.org/10.1103/PhysRevD.72.074004} {\bibfield  {journal} {\bibinfo  {journal} {Phys. Rev. D}\ }\textbf {\bibinfo {volume} {72}},\ \bibinfo {pages} {074004} (\bibinfo {year} {2005})},\ \Eprint {https://arxiv.org/abs/hep-ph/0505195} {arXiv:hep-ph/0505195} \BibitemShut {NoStop}%
\bibitem [{\citenamefont {Close}\ and\ \citenamefont {Swanson}(2005)}]{Close:2005se}%
  \BibitemOpen
  \bibfield  {author} {\bibinfo {author} {\bibfnamefont {F.~E.}\ \bibnamefont {Close}}\ and\ \bibinfo {author} {\bibfnamefont {E.~S.}\ \bibnamefont {Swanson}},\ }\bibinfo {title} {{Dynamics and decay of Heavy-Light hadrons}},\ \href {https://doi.org/10.1103/PhysRevD.72.094004} {\bibfield  {journal} {\bibinfo  {journal} {Phys. Rev. D}\ }\textbf {\bibinfo {volume} {72}},\ \bibinfo {pages} {094004} (\bibinfo {year} {2005})},\ \Eprint {https://arxiv.org/abs/hep-ph/0505206} {arXiv:hep-ph/0505206} \BibitemShut {NoStop}%
\bibitem [{\citenamefont {Liu}\ {\it et~al.}(2006)\citenamefont {Liu}, \citenamefont {Yu}, \citenamefont {Zhao},\ and\ \citenamefont {Li}}]{Liu:2006jx}%
  \BibitemOpen
  \bibfield  {author} {\bibinfo {author} {\bibfnamefont {X.}~\bibnamefont {Liu}}, \bibinfo {author} {\bibfnamefont {Y.-M.}\ \bibnamefont {Yu}}, \bibinfo {author} {\bibfnamefont {S.-M.}\ \bibnamefont {Zhao}},\ and\ \bibinfo {author} {\bibfnamefont {X.-Q.}\ \bibnamefont {Li}},\ }\bibinfo {title} {{Study on decays of $D^{*}_{sJ}(2317)$ and $D_{sJ}(2460)$ in terms of the CQM model}},\ \href {https://doi.org/10.1140/epjc/s2006-02564-0} {\bibfield  {journal} {\bibinfo  {journal} {Eur. Phys. J. C}\ }\textbf {\bibinfo {volume} {47}},\ \bibinfo {pages} {445} (\bibinfo {year} {2006})},\ \Eprint {https://arxiv.org/abs/hep-ph/0601017} {arXiv:hep-ph/0601017} \BibitemShut {NoStop}%
\bibitem [{\citenamefont {Wang}(2007{\natexlab{b}})}]{Wang:2006zw}%
  \BibitemOpen
  \bibfield  {author} {\bibinfo {author} {\bibfnamefont {Z.~G.}\ \bibnamefont {Wang}},\ }\bibinfo {title} {{Structure of the axial-vector meson $D_{s1}(2460)$}},\ \href {https://doi.org/10.1088/0954-3899/34/4/011} {\bibfield  {journal} {\bibinfo  {journal} {J. Phys. G}\ }\textbf {\bibinfo {volume} {34}},\ \bibinfo {pages} {753} (\bibinfo {year} {2007}{\natexlab{b}})},\ \Eprint {https://arxiv.org/abs/hep-ph/0611271} {arXiv:hep-ph/0611271} \BibitemShut {NoStop}%
\bibitem [{\citenamefont {van Beveren}\ and\ \citenamefont {Rupp}(2003)}]{vanBeveren:2003kd}%
  \BibitemOpen
  \bibfield  {author} {\bibinfo {author} {\bibfnamefont {E.}~\bibnamefont {van Beveren}}\ and\ \bibinfo {author} {\bibfnamefont {G.}~\bibnamefont {Rupp}},\ }\bibinfo {title} {{Observed $D_s(2317)$ and tentative $D(2100\text{--}2300)$ as the charmed cousins of the light scalar nonet}},\ \href {https://doi.org/10.1103/PhysRevLett.91.012003} {\bibfield  {journal} {\bibinfo  {journal} {Phys. Rev. Lett.}\ }\textbf {\bibinfo {volume} {91}},\ \bibinfo {pages} {012003} (\bibinfo {year} {2003})},\ \Eprint {https://arxiv.org/abs/hep-ph/0305035} {arXiv:hep-ph/0305035} \BibitemShut {NoStop}%
\bibitem [{\citenamefont {Barnes}\ {\it et~al.}(2003)\citenamefont {Barnes}, \citenamefont {Close},\ and\ \citenamefont {Lipkin}}]{Barnes:2003dj}%
  \BibitemOpen
  \bibfield  {author} {\bibinfo {author} {\bibfnamefont {T.}~\bibnamefont {Barnes}}, \bibinfo {author} {\bibfnamefont {F.~E.}\ \bibnamefont {Close}},\ and\ \bibinfo {author} {\bibfnamefont {H.~J.}\ \bibnamefont {Lipkin}},\ }\bibinfo {title} {{Implications of a DK molecule at 2.32 $\mathrm{GeV}$}},\ \href {https://doi.org/10.1103/PhysRevD.68.054006} {\bibfield  {journal} {\bibinfo  {journal} {Phys. Rev. D}\ }\textbf {\bibinfo {volume} {68}},\ \bibinfo {pages} {054006} (\bibinfo {year} {2003})},\ \Eprint {https://arxiv.org/abs/hep-ph/0305025} {arXiv:hep-ph/0305025} \BibitemShut {NoStop}%
\bibitem [{\citenamefont {Chen}\ and\ \citenamefont {Li}(2004)}]{Chen:2004dy}%
  \BibitemOpen
  \bibfield  {author} {\bibinfo {author} {\bibfnamefont {Y.-Q.}\ \bibnamefont {Chen}}\ and\ \bibinfo {author} {\bibfnamefont {X.-Q.}\ \bibnamefont {Li}},\ }\bibinfo {title} {{A Comprehensive Four-Quark Interpretation of $D_{s}(2317)$, $D_{s}(2457)$, and $D_{s}(2632)$}},\ \href {https://doi.org/10.1103/PhysRevLett.93.232001} {\bibfield  {journal} {\bibinfo  {journal} {Phys. Rev. Lett.}\ }\textbf {\bibinfo {volume} {93}},\ \bibinfo {pages} {232001} (\bibinfo {year} {2004})},\ \Eprint {https://arxiv.org/abs/hep-ph/0407062} {arXiv:hep-ph/0407062} \BibitemShut {NoStop}%
\bibitem [{\citenamefont {Kolomeitsev}\ and\ \citenamefont {Lutz}(2004)}]{Kolomeitsev:2003ac}%
  \BibitemOpen
  \bibfield  {author} {\bibinfo {author} {\bibfnamefont {E.~E.}\ \bibnamefont {Kolomeitsev}}\ and\ \bibinfo {author} {\bibfnamefont {M.~F.~M.}\ \bibnamefont {Lutz}},\ }\bibinfo {title} {{On heavy-light meson resonances and chiral symmetry}},\ \href {https://doi.org/10.1016/j.physletb.2003.10.118} {\bibfield  {journal} {\bibinfo  {journal} {Phys. Lett. B}\ }\textbf {\bibinfo {volume} {582}},\ \bibinfo {pages} {39} (\bibinfo {year} {2004})},\ \Eprint {https://arxiv.org/abs/hep-ph/0307133} {arXiv:hep-ph/0307133} \BibitemShut {NoStop}%
\bibitem [{\citenamefont {Gamermann}\ {\it et~al.}(2007{\natexlab{a}})\citenamefont {Gamermann}, \citenamefont {Oset}, \citenamefont {Strottman},\ and\ \citenamefont {Vicente~Vacas}}]{Gamermann:2006nm}%
  \BibitemOpen
  \bibfield  {author} {\bibinfo {author} {\bibfnamefont {D.}~\bibnamefont {Gamermann}}, \bibinfo {author} {\bibfnamefont {E.}~\bibnamefont {Oset}}, \bibinfo {author} {\bibfnamefont {D.}~\bibnamefont {Strottman}},\ and\ \bibinfo {author} {\bibfnamefont {M.~J.}\ \bibnamefont {Vicente~Vacas}},\ }\bibinfo {title} {{Dynamically generated open and hidden charm meson systems}},\ \href {https://doi.org/10.1103/PhysRevD.76.074016} {\bibfield  {journal} {\bibinfo  {journal} {Phys. Rev. D}\ }\textbf {\bibinfo {volume} {76}},\ \bibinfo {pages} {074016} (\bibinfo {year} {2007}{\natexlab{a}})},\ \Eprint {https://arxiv.org/abs/hep-ph/0612179} {arXiv:hep-ph/0612179} \BibitemShut {NoStop}%
\bibitem [{\citenamefont {Guo}\ {\it et~al.}(2007)\citenamefont {Guo}, \citenamefont {Shen},\ and\ \citenamefont {Chiang}}]{Guo:2006rp}%
  \BibitemOpen
  \bibfield  {author} {\bibinfo {author} {\bibfnamefont {F.-K.}\ \bibnamefont {Guo}}, \bibinfo {author} {\bibfnamefont {P.-N.}\ \bibnamefont {Shen}},\ and\ \bibinfo {author} {\bibfnamefont {H.-C.}\ \bibnamefont {Chiang}},\ }\bibinfo {title} {{Dynamically generated $1^{+}$ heavy mesons}},\ \href {https://doi.org/10.1016/j.physletb.2007.01.050} {\bibfield  {journal} {\bibinfo  {journal} {Phys. Lett. B}\ }\textbf {\bibinfo {volume} {647}},\ \bibinfo {pages} {133} (\bibinfo {year} {2007})},\ \Eprint {https://arxiv.org/abs/hep-ph/0610008} {arXiv:hep-ph/0610008} \BibitemShut {NoStop}%
\bibitem [{\citenamefont {Yang}\ {\it et~al.}(2022)\citenamefont {Yang}, \citenamefont {Wang}, \citenamefont {Wu}, \citenamefont {Oka},\ and\ \citenamefont {Zhu}}]{Yang:2021tvc}%
  \BibitemOpen
  \bibfield  {author} {\bibinfo {author} {\bibfnamefont {Z.}~\bibnamefont {Yang}}, \bibinfo {author} {\bibfnamefont {G.-J.}\ \bibnamefont {Wang}}, \bibinfo {author} {\bibfnamefont {J.-J.}\ \bibnamefont {Wu}}, \bibinfo {author} {\bibfnamefont {M.}~\bibnamefont {Oka}},\ and\ \bibinfo {author} {\bibfnamefont {S.-L.}\ \bibnamefont {Zhu}},\ }\bibinfo {title} {{Novel Coupled Channel Framework Connecting the Quark Model and Lattice QCD for the Near-threshold $D_{s}$ States}},\ \href {https://doi.org/10.1103/PhysRevLett.128.112001} {\bibfield  {journal} {\bibinfo  {journal} {Phys. Rev. Lett.}\ }\textbf {\bibinfo {volume} {128}},\ \bibinfo {pages} {112001} (\bibinfo {year} {2022})},\ \Eprint {https://arxiv.org/abs/2107.04860} {arXiv:2107.04860 [hep-ph]} \BibitemShut {NoStop}%
\bibitem [{\citenamefont {Liu}\ {\it et~al.}(2022)\citenamefont {Liu}, \citenamefont {Ling}, \citenamefont {Geng}, \citenamefont {En-Wang},\ and\ \citenamefont {Xie}}]{Liu:2022dmm}%
  \BibitemOpen
  \bibfield  {author} {\bibinfo {author} {\bibfnamefont {M.-Z.}\ \bibnamefont {Liu}}, \bibinfo {author} {\bibfnamefont {X.-Z.}\ \bibnamefont {Ling}}, \bibinfo {author} {\bibfnamefont {L.-S.}\ \bibnamefont {Geng}}, \bibinfo {author} {\bibnamefont {En-Wang}},\ and\ \bibinfo {author} {\bibfnamefont {J.-J.}\ \bibnamefont {Xie}},\ }\bibinfo {title} {{Production of $D_{s0}^{*}(2317)$ and $D_{s1}(2460)$ in $B$ decays as $D^{(*)}K$ and $D_{s}^{(*)}\eta$ molecules}},\ \href {https://doi.org/10.1103/PhysRevD.106.114011} {\bibfield  {journal} {\bibinfo  {journal} {Phys. Rev. D}\ }\textbf {\bibinfo {volume} {106}},\ \bibinfo {pages} {114011} (\bibinfo {year} {2022})},\ \Eprint {https://arxiv.org/abs/2209.01103} {arXiv:2209.01103 [hep-ph]} \BibitemShut {NoStop}%
\bibitem [{\citenamefont {Mohler}\ {\it et~al.}(2013)\citenamefont {Mohler}, \citenamefont {Lang}, \citenamefont {Leskovec}, \citenamefont {Prelovsek},\ and\ \citenamefont {Woloshyn}}]{Mohler:2013rwa}%
  \BibitemOpen
  \bibfield  {author} {\bibinfo {author} {\bibfnamefont {D.}~\bibnamefont {Mohler}}, \bibinfo {author} {\bibfnamefont {C.~B.}\ \bibnamefont {Lang}}, \bibinfo {author} {\bibfnamefont {L.}~\bibnamefont {Leskovec}}, \bibinfo {author} {\bibfnamefont {S.}~\bibnamefont {Prelovsek}},\ and\ \bibinfo {author} {\bibfnamefont {R.~M.}\ \bibnamefont {Woloshyn}},\ }\bibinfo {title} {{$D_{s0}^*(2317)$ Meson and $DK$ Scattering from Lattice QCD}},\ \href {https://doi.org/10.1103/PhysRevLett.111.222001} {\bibfield  {journal} {\bibinfo  {journal} {Phys. Rev. Lett.}\ }\textbf {\bibinfo {volume} {111}},\ \bibinfo {pages} {222001} (\bibinfo {year} {2013})},\ \Eprint {https://arxiv.org/abs/1308.3175} {arXiv:1308.3175 [hep-lat]} \BibitemShut {NoStop}%
\bibitem [{\citenamefont {Lang}\ {\it et~al.}(2014)\citenamefont {Lang}, \citenamefont {Leskovec}, \citenamefont {Mohler}, \citenamefont {Prelovsek},\ and\ \citenamefont {Woloshyn}}]{Lang:2014yfa}%
  \BibitemOpen
  \bibfield  {author} {\bibinfo {author} {\bibfnamefont {C.~B.}\ \bibnamefont {Lang}}, \bibinfo {author} {\bibfnamefont {L.}~\bibnamefont {Leskovec}}, \bibinfo {author} {\bibfnamefont {D.}~\bibnamefont {Mohler}}, \bibinfo {author} {\bibfnamefont {S.}~\bibnamefont {Prelovsek}},\ and\ \bibinfo {author} {\bibfnamefont {R.~M.}\ \bibnamefont {Woloshyn}},\ }\bibinfo {title} {{$D_s$ Mesons with $DK$ and $D^{*}K$ Scattering Near Threshold}},\ \href {https://doi.org/10.1103/PhysRevD.90.034510} {\bibfield  {journal} {\bibinfo  {journal} {Phys. Rev. D}\ }\textbf {\bibinfo {volume} {90}},\ \bibinfo {pages} {034510} (\bibinfo {year} {2014})},\ \Eprint {https://arxiv.org/abs/1403.8103} {arXiv:1403.8103 [hep-lat]} \BibitemShut {NoStop}%
\bibitem [{\citenamefont {Bali}\ {\it et~al.}(2017)\citenamefont {Bali}, \citenamefont {Collins}, \citenamefont {Cox},\ and\ \citenamefont {Sch{\"a}fer}}]{Bali:2017pdv}%
  \BibitemOpen
  \bibfield  {author} {\bibinfo {author} {\bibfnamefont {G.~S.}\ \bibnamefont {Bali}}, \bibinfo {author} {\bibfnamefont {S.}~\bibnamefont {Collins}}, \bibinfo {author} {\bibfnamefont {A.}~\bibnamefont {Cox}},\ and\ \bibinfo {author} {\bibfnamefont {A.}~\bibnamefont {Sch{\"a}fer}},\ }\bibinfo {title} {{Masses and decay constants of the $D_{s0}^*(2317)$ and $D_{s1}(2460)$ from $N_f=2$ lattice QCD close to the physical point}},\ \href {https://doi.org/10.1103/PhysRevD.96.074501} {\bibfield  {journal} {\bibinfo  {journal} {Phys. Rev. D}\ }\textbf {\bibinfo {volume} {96}},\ \bibinfo {pages} {074501} (\bibinfo {year} {2017})},\ \Eprint {https://arxiv.org/abs/1706.01247} {arXiv:1706.01247 [hep-lat]} \BibitemShut {NoStop}%
\bibitem [{\citenamefont {Cheung}\ {\it et~al.}(2021)\citenamefont {Cheung}, \citenamefont {Thomas}, \citenamefont {Wilson}, \citenamefont {Moir}, \citenamefont {Peardon},\ and\ \citenamefont {Ryan}}]{Cheung:2020mql}%
  \BibitemOpen
  \bibfield  {author} {\bibinfo {author} {\bibfnamefont {G.~K.~C.}\ \bibnamefont {Cheung}}, \bibinfo {author} {\bibfnamefont {C.~E.}\ \bibnamefont {Thomas}}, \bibinfo {author} {\bibfnamefont {D.~J.}\ \bibnamefont {Wilson}}, \bibinfo {author} {\bibfnamefont {G.}~\bibnamefont {Moir}}, \bibinfo {author} {\bibfnamefont {M.}~\bibnamefont {Peardon}},\ and\ \bibinfo {author} {\bibfnamefont {S.~M.}\ \bibnamefont {Ryan}} (\bibinfo {collaboration} {Hadron Spectrum}),\ }\bibinfo {title} {{$DK~ I = 0$,$ D\bar{K}~I = 0$, 1 scattering and the $ {D}_{s0}^{\ast } $(2317) from lattice QCD}},\ \href {https://doi.org/10.1007/JHEP02(2021)100} {\bibfield  {journal} {\bibinfo  {journal} {JHEP}\ }\textbf {\bibinfo {volume} {02}},\ \bibinfo {pages} {100}},\ \Eprint {https://arxiv.org/abs/2008.06432} {arXiv:2008.06432 [hep-lat]} \BibitemShut {NoStop}%
\bibitem [{\citenamefont {Mart{\'\i}nez~Torres}\ {\it et~al.}(2015)\citenamefont {Mart{\'\i}nez~Torres}, \citenamefont {Oset}, \citenamefont {Prelovsek},\ and\ \citenamefont {Ramos}}]{MartinezTorres:2014kpc}%
  \BibitemOpen
  \bibfield  {author} {\bibinfo {author} {\bibfnamefont {A.}~\bibnamefont {Mart{\'\i}nez~Torres}}, \bibinfo {author} {\bibfnamefont {E.}~\bibnamefont {Oset}}, \bibinfo {author} {\bibfnamefont {S.}~\bibnamefont {Prelovsek}},\ and\ \bibinfo {author} {\bibfnamefont {A.}~\bibnamefont {Ramos}},\ }\bibinfo {title} {{Reanalysis of lattice QCD spectra leading to the $D_{s0}^*(2317)$ and $D_{s1}^*(2460)$}},\ \href {https://doi.org/10.1007/JHEP05(2015)153} {\bibfield  {journal} {\bibinfo  {journal} {JHEP}\ }\textbf {\bibinfo {volume} {05}},\ \bibinfo {pages} {153}},\ \Eprint {https://arxiv.org/abs/1412.1706} {arXiv:1412.1706 [hep-lat]} \BibitemShut {NoStop}%
\bibitem [{\citenamefont {Faessler}\ {\it et~al.}(2007)\citenamefont {Faessler}, \citenamefont {Gutsche}, \citenamefont {Lyubovitskij},\ and\ \citenamefont {Ma}}]{Faessler:2007gv}%
  \BibitemOpen
  \bibfield  {author} {\bibinfo {author} {\bibfnamefont {A.}~\bibnamefont {Faessler}}, \bibinfo {author} {\bibfnamefont {T.}~\bibnamefont {Gutsche}}, \bibinfo {author} {\bibfnamefont {V.~E.}\ \bibnamefont {Lyubovitskij}},\ and\ \bibinfo {author} {\bibfnamefont {Y.-L.}\ \bibnamefont {Ma}},\ }\bibinfo {title} {{Strong and radiative decays of the $D_{s0}^{*}(2317)$ meson in the $DK$-molecule picture}},\ \href {https://doi.org/10.1103/PhysRevD.76.014005} {\bibfield  {journal} {\bibinfo  {journal} {Phys. Rev. D}\ }\textbf {\bibinfo {volume} {76}},\ \bibinfo {pages} {014005} (\bibinfo {year} {2007})},\ \Eprint {https://arxiv.org/abs/0705.0254} {arXiv:0705.0254 [hep-ph]} \BibitemShut {NoStop}%
\bibitem [{\citenamefont {Gamermann}\ {\it et~al.}(2007{\natexlab{b}})\citenamefont {Gamermann}, \citenamefont {Dai},\ and\ \citenamefont {Oset}}]{Gamermann:2007bm}%
  \BibitemOpen
  \bibfield  {author} {\bibinfo {author} {\bibfnamefont {D.}~\bibnamefont {Gamermann}}, \bibinfo {author} {\bibfnamefont {L.~R.}\ \bibnamefont {Dai}},\ and\ \bibinfo {author} {\bibfnamefont {E.}~\bibnamefont {Oset}},\ }\bibinfo {title} {{Radiative decay of the dynamically generated open and hidden charm scalar meson resonances $D_{s0}^{*}.(2317)$ and $X(3700)$}},\ \href {https://doi.org/10.1103/PhysRevC.76.055205} {\bibfield  {journal} {\bibinfo  {journal} {Phys. Rev. C}\ }\textbf {\bibinfo {volume} {76}},\ \bibinfo {pages} {055205} (\bibinfo {year} {2007}{\natexlab{b}})},\ \Eprint {https://arxiv.org/abs/0709.2339} {arXiv:0709.2339 [hep-ph]} \BibitemShut {NoStop}%
\bibitem [{\citenamefont {Lutz}\ and\ \citenamefont {Soyeur}(2008)}]{Lutz:2007sk}%
  \BibitemOpen
  \bibfield  {author} {\bibinfo {author} {\bibfnamefont {M.~F.~M.}\ \bibnamefont {Lutz}}\ and\ \bibinfo {author} {\bibfnamefont {M.}~\bibnamefont {Soyeur}},\ }\bibinfo {title} {{Radiative and isospin-violating decays of $D_{s}$-mesons in the hadrogenesis conjecture}},\ \href {https://doi.org/10.1016/j.nuclphysa.2008.09.003} {\bibfield  {journal} {\bibinfo  {journal} {Nucl. Phys. A}\ }\textbf {\bibinfo {volume} {813}},\ \bibinfo {pages} {14} (\bibinfo {year} {2008})},\ \Eprint {https://arxiv.org/abs/0710.1545} {arXiv:0710.1545 [hep-ph]} \BibitemShut {NoStop}%
\bibitem [{\citenamefont {Cleven}\ {\it et~al.}(2014)\citenamefont {Cleven}, \citenamefont {Grie{\ss}hammer}, \citenamefont {Guo}, \citenamefont {Hanhart},\ and\ \citenamefont {Mei{\ss}ner}}]{Cleven:2014oka}%
  \BibitemOpen
  \bibfield  {author} {\bibinfo {author} {\bibfnamefont {M.}~\bibnamefont {Cleven}}, \bibinfo {author} {\bibfnamefont {H.~W.}\ \bibnamefont {Grie{\ss}hammer}}, \bibinfo {author} {\bibfnamefont {F.-K.}\ \bibnamefont {Guo}}, \bibinfo {author} {\bibfnamefont {C.}~\bibnamefont {Hanhart}},\ and\ \bibinfo {author} {\bibfnamefont {U.-G.}\ \bibnamefont {Mei{\ss}ner}},\ }\bibinfo {title} {{Strong and radiative decays of the $D^*_{s0}(2317)$ and $D_{s1}(2460)$}},\ \href {https://doi.org/10.1140/epja/i2014-14149-y} {\bibfield  {journal} {\bibinfo  {journal} {Eur. Phys. J. A}\ }\textbf {\bibinfo {volume} {50}},\ \bibinfo {pages} {149} (\bibinfo {year} {2014})},\ \Eprint {https://arxiv.org/abs/1405.2242} {arXiv:1405.2242 [hep-ph]} \BibitemShut {NoStop}%
\bibitem [{\citenamefont {Fu}\ {\it et~al.}(2022)\citenamefont {Fu}, \citenamefont {Grie{\ss}hammer}, \citenamefont {Guo}, \citenamefont {Hanhart},\ and\ \citenamefont {Mei{\ss}ner}}]{Fu:2021wde}%
  \BibitemOpen
  \bibfield  {author} {\bibinfo {author} {\bibfnamefont {H.-L.}\ \bibnamefont {Fu}}, \bibinfo {author} {\bibfnamefont {H.~W.}\ \bibnamefont {Grie{\ss}hammer}}, \bibinfo {author} {\bibfnamefont {F.-K.}\ \bibnamefont {Guo}}, \bibinfo {author} {\bibfnamefont {C.}~\bibnamefont {Hanhart}},\ and\ \bibinfo {author} {\bibfnamefont {U.-G.}\ \bibnamefont {Mei{\ss}ner}},\ }\bibinfo {title} {{Update on strong and radiative decays of the $D_{s0}^*(2317)$ and $D_{s1}(2460)$ and their bottom cousins}},\ \href {https://doi.org/10.1140/epja/s10050-022-00724-8} {\bibfield  {journal} {\bibinfo  {journal} {Eur. Phys. J. A}\ }\textbf {\bibinfo {volume} {58}},\ \bibinfo {pages} {70} (\bibinfo {year} {2022})},\ \Eprint {https://arxiv.org/abs/2111.09481} {arXiv:2111.09481 [hep-ph]} \BibitemShut {NoStop}%
\bibitem [{\citenamefont {Abumusabh}\ {\it et~al.}(2025)\citenamefont {Abumusabh} {\it et~al.}}]{Belle-II:2025dzk}%
  \BibitemOpen
  \bibfield  {author} {\bibinfo {author} {\bibfnamefont {M.}~\bibnamefont {Abumusabh}} {\it et~al.} (\bibinfo {collaboration} {Belle-II}),\ }\href@noop {} {\bibinfo {title} {{Observation of the radiative decay $D_s(2317)^+ \to D_s^* \gamma$}}} (\bibinfo {year} {2025}),\ \Eprint {https://arxiv.org/abs/2510.27174} {arXiv:2510.27174 [hep-ex]} \BibitemShut {NoStop}%
\bibitem [{\citenamefont {Yao}\ {\it et~al.}(2006)\citenamefont {Yao} {\it et~al.}}]{ParticleDataGroup:2006fqo}%
  \BibitemOpen
  \bibfield  {author} {\bibinfo {author} {\bibfnamefont {W.~M.}\ \bibnamefont {Yao}} {\it et~al.} (\bibinfo {collaboration} {Particle Data Group}),\ }\bibinfo {title} {{Review of Particle Physics}},\ \href {https://doi.org/10.1088/0954-3899/33/1/001} {\bibfield  {journal} {\bibinfo  {journal} {J. Phys. G}\ }\textbf {\bibinfo {volume} {33}},\ \bibinfo {pages} {1} (\bibinfo {year} {2006})}\BibitemShut {NoStop}%
\bibitem [{\citenamefont {Bando}\ {\it et~al.}(1985)\citenamefont {Bando}, \citenamefont {Kugo}, \citenamefont {Uehara}, \citenamefont {Yamawaki},\ and\ \citenamefont {Yanagida}}]{Bando:1984ej}%
  \BibitemOpen
  \bibfield  {author} {\bibinfo {author} {\bibfnamefont {M.}~\bibnamefont {Bando}}, \bibinfo {author} {\bibfnamefont {T.}~\bibnamefont {Kugo}}, \bibinfo {author} {\bibfnamefont {S.}~\bibnamefont {Uehara}}, \bibinfo {author} {\bibfnamefont {K.}~\bibnamefont {Yamawaki}},\ and\ \bibinfo {author} {\bibfnamefont {T.}~\bibnamefont {Yanagida}},\ }\bibinfo {title} {{Is the $\rho$ Meson a Dynamical Gauge Boson of Hidden Local Symmetry?}},\ \href {https://doi.org/10.1103/PhysRevLett.54.1215} {\bibfield  {journal} {\bibinfo  {journal} {Phys. Rev. Lett.}\ }\textbf {\bibinfo {volume} {54}},\ \bibinfo {pages} {1215} (\bibinfo {year} {1985})}\BibitemShut {NoStop}%
\bibitem [{\citenamefont {Bando}\ {\it et~al.}(1988)\citenamefont {Bando}, \citenamefont {Kugo},\ and\ \citenamefont {Yamawaki}}]{Bando:1987br}%
  \BibitemOpen
  \bibfield  {author} {\bibinfo {author} {\bibfnamefont {M.}~\bibnamefont {Bando}}, \bibinfo {author} {\bibfnamefont {T.}~\bibnamefont {Kugo}},\ and\ \bibinfo {author} {\bibfnamefont {K.}~\bibnamefont {Yamawaki}},\ }\bibinfo {title} {{Nonlinear Realization and Hidden Local Symmetries}},\ \href {https://doi.org/10.1016/0370-1573(88)90019-1} {\bibfield  {journal} {\bibinfo  {journal} {Phys. Rept.}\ }\textbf {\bibinfo {volume} {164}},\ \bibinfo {pages} {217} (\bibinfo {year} {1988})}\BibitemShut {NoStop}%
\bibitem [{\citenamefont {Meissner}(1988)}]{Meissner:1987ge}%
  \BibitemOpen
  \bibfield  {author} {\bibinfo {author} {\bibfnamefont {U.~G.}\ \bibnamefont {Meissner}},\ }\bibinfo {title} {{Low-Energy Hadron Physics from Effective Chiral Lagrangians with Vector Mesons}},\ \href {https://doi.org/10.1016/0370-1573(88)90090-7} {\bibfield  {journal} {\bibinfo  {journal} {Phys. Rept.}\ }\textbf {\bibinfo {volume} {161}},\ \bibinfo {pages} {213} (\bibinfo {year} {1988})}\BibitemShut {NoStop}%
\bibitem [{\citenamefont {Nagahiro}\ {\it et~al.}(2009)\citenamefont {Nagahiro}, \citenamefont {Roca}, \citenamefont {Hosaka},\ and\ \citenamefont {Oset}}]{Nagahiro:2008cv}%
  \BibitemOpen
  \bibfield  {author} {\bibinfo {author} {\bibfnamefont {H.}~\bibnamefont {Nagahiro}}, \bibinfo {author} {\bibfnamefont {L.}~\bibnamefont {Roca}}, \bibinfo {author} {\bibfnamefont {A.}~\bibnamefont {Hosaka}},\ and\ \bibinfo {author} {\bibfnamefont {E.}~\bibnamefont {Oset}},\ }\bibinfo {title} {{Hidden gauge formalism for the radiative decays of axial-vector mesons}},\ \href {https://doi.org/10.1103/PhysRevD.79.014015} {\bibfield  {journal} {\bibinfo  {journal} {Phys. Rev. D}\ }\textbf {\bibinfo {volume} {79}},\ \bibinfo {pages} {014015} (\bibinfo {year} {2009})},\ \Eprint {https://arxiv.org/abs/0809.0943} {arXiv:0809.0943 [hep-ph]} \BibitemShut {NoStop}%
\bibitem [{\citenamefont {Ikeno}\ {\it et~al.}(2023)\citenamefont {Ikeno}, \citenamefont {Toledo},\ and\ \citenamefont {Oset}}]{Ikeno:2023ojl}%
  \BibitemOpen
  \bibfield  {author} {\bibinfo {author} {\bibfnamefont {N.}~\bibnamefont {Ikeno}}, \bibinfo {author} {\bibfnamefont {G.}~\bibnamefont {Toledo}},\ and\ \bibinfo {author} {\bibfnamefont {E.}~\bibnamefont {Oset}},\ }\bibinfo {title} {{Model independent analysis of femtoscopic correlation functions: An application to the $D_{s0}^{*}$(2317)}},\ \href {https://doi.org/10.1016/j.physletb.2023.138281} {\bibfield  {journal} {\bibinfo  {journal} {Phys. Lett. B}\ }\textbf {\bibinfo {volume} {847}},\ \bibinfo {pages} {138281} (\bibinfo {year} {2023})},\ \Eprint {https://arxiv.org/abs/2305.16431} {arXiv:2305.16431 [hep-ph]} \BibitemShut {NoStop}%
\bibitem [{\citenamefont {Ablikim}\ {\it et~al.}(2018)\citenamefont {Ablikim} {\it et~al.}}]{BESIII:2017vdm}%
  \BibitemOpen
  \bibfield  {author} {\bibinfo {author} {\bibfnamefont {M.}~\bibnamefont {Ablikim}} {\it et~al.} (\bibinfo {collaboration} {BESIII}),\ }\bibinfo {title} {{Measurement of the absolute branching fraction of $D_{s0}^{*\pm}(2317)\rightarrow \pi^0 D_{s}^{\pm}$}},\ \href {https://doi.org/10.1103/PhysRevD.97.051103} {\bibfield  {journal} {\bibinfo  {journal} {Phys. Rev. D}\ }\textbf {\bibinfo {volume} {97}},\ \bibinfo {pages} {051103} (\bibinfo {year} {2018})},\ \Eprint {https://arxiv.org/abs/1711.08293} {arXiv:1711.08293 [hep-ex]} \BibitemShut {NoStop}%
\bibitem [{\citenamefont {Guo}\ {\it et~al.}(2006)\citenamefont {Guo}, \citenamefont {Shen}, \citenamefont {Chiang}, \citenamefont {Ping},\ and\ \citenamefont {Zou}}]{Guo:2006fu}%
  \BibitemOpen
  \bibfield  {author} {\bibinfo {author} {\bibfnamefont {F.-K.}\ \bibnamefont {Guo}}, \bibinfo {author} {\bibfnamefont {P.-N.}\ \bibnamefont {Shen}}, \bibinfo {author} {\bibfnamefont {H.-C.}\ \bibnamefont {Chiang}}, \bibinfo {author} {\bibfnamefont {R.-G.}\ \bibnamefont {Ping}},\ and\ \bibinfo {author} {\bibfnamefont {B.-S.}\ \bibnamefont {Zou}},\ }\bibinfo {title} {{Dynamically generated $0^{+}$ heavy mesons in a heavy chiral unitary approach}},\ \href {https://doi.org/10.1016/j.physletb.2006.08.064} {\bibfield  {journal} {\bibinfo  {journal} {Phys. Lett. B}\ }\textbf {\bibinfo {volume} {641}},\ \bibinfo {pages} {278} (\bibinfo {year} {2006})},\ \Eprint {https://arxiv.org/abs/hep-ph/0603072} {arXiv:hep-ph/0603072} \BibitemShut {NoStop}%
\bibitem [{\citenamefont {Guo}\ {\it et~al.}(2008)\citenamefont {Guo}, \citenamefont {Hanhart}, \citenamefont {Krewald},\ and\ \citenamefont {Meissner}}]{Guo:2008gp}%
  \BibitemOpen
  \bibfield  {author} {\bibinfo {author} {\bibfnamefont {F.-K.}\ \bibnamefont {Guo}}, \bibinfo {author} {\bibfnamefont {C.}~\bibnamefont {Hanhart}}, \bibinfo {author} {\bibfnamefont {S.}~\bibnamefont {Krewald}},\ and\ \bibinfo {author} {\bibfnamefont {U.-G.}\ \bibnamefont {Meissner}},\ }\bibinfo {title} {{Subleading contributions to the width of the $D^{*}_{s0}(2317)$}},\ \href {https://doi.org/10.1016/j.physletb.2008.07.060} {\bibfield  {journal} {\bibinfo  {journal} {Phys. Lett. B}\ }\textbf {\bibinfo {volume} {666}},\ \bibinfo {pages} {251} (\bibinfo {year} {2008})},\ \Eprint {https://arxiv.org/abs/0806.3374} {arXiv:0806.3374 [hep-ph]} \BibitemShut {NoStop}%
\bibitem [{\citenamefont {Liu}\ {\it et~al.}(2013)\citenamefont {Liu}, \citenamefont {Orginos}, \citenamefont {Guo}, \citenamefont {Hanhart},\ and\ \citenamefont {Meissner}}]{Liu:2012zya}%
  \BibitemOpen
  \bibfield  {author} {\bibinfo {author} {\bibfnamefont {L.}~\bibnamefont {Liu}}, \bibinfo {author} {\bibfnamefont {K.}~\bibnamefont {Orginos}}, \bibinfo {author} {\bibfnamefont {F.-K.}\ \bibnamefont {Guo}}, \bibinfo {author} {\bibfnamefont {C.}~\bibnamefont {Hanhart}},\ and\ \bibinfo {author} {\bibfnamefont {U.-G.}\ \bibnamefont {Meissner}},\ }\bibinfo {title} {{Interactions of charmed mesons with light pseudoscalar mesons from lattice QCD and implications on the nature of the $D_{s0}^*(2317)$}},\ \href {https://doi.org/10.1103/PhysRevD.87.014508} {\bibfield  {journal} {\bibinfo  {journal} {Phys. Rev. D}\ }\textbf {\bibinfo {volume} {87}},\ \bibinfo {pages} {014508} (\bibinfo {year} {2013})},\ \Eprint {https://arxiv.org/abs/1208.4535} {arXiv:1208.4535 [hep-lat]} \BibitemShut {NoStop}%
\bibitem [{\citenamefont {Fajfer}\ and\ \citenamefont {Prapotnik~Brdnik}(2015)}]{Fajfer:2015zma}%
  \BibitemOpen
  \bibfield  {author} {\bibinfo {author} {\bibfnamefont {S.}~\bibnamefont {Fajfer}}\ and\ \bibinfo {author} {\bibfnamefont {A.}~\bibnamefont {Prapotnik~Brdnik}},\ }\bibinfo {title} {{Chiral loops in the isospin violating decays of $D_{s1}(2460)^+$ and $D*_{s0}(2317)^+$}},\ \href {https://doi.org/10.1103/PhysRevD.92.074047} {\bibfield  {journal} {\bibinfo  {journal} {Phys. Rev. D}\ }\textbf {\bibinfo {volume} {92}},\ \bibinfo {pages} {074047} (\bibinfo {year} {2015})},\ \Eprint {https://arxiv.org/abs/1506.02716} {arXiv:1506.02716 [hep-ph]} \BibitemShut {NoStop}%
\bibitem [{\citenamefont {Guo}\ {\it et~al.}(2018)\citenamefont {Guo}, \citenamefont {Heo},\ and\ \citenamefont {Lutz}}]{Guo:2018kno}%
  \BibitemOpen
  \bibfield  {author} {\bibinfo {author} {\bibfnamefont {X.-Y.}\ \bibnamefont {Guo}}, \bibinfo {author} {\bibfnamefont {Y.}~\bibnamefont {Heo}},\ and\ \bibinfo {author} {\bibfnamefont {M.~F.~M.}\ \bibnamefont {Lutz}},\ }\bibinfo {title} {{On chiral extrapolations of charmed meson masses and coupled-channel reaction dynamics}},\ \href {https://doi.org/10.1103/PhysRevD.98.014510} {\bibfield  {journal} {\bibinfo  {journal} {Phys. Rev. D}\ }\textbf {\bibinfo {volume} {98}},\ \bibinfo {pages} {014510} (\bibinfo {year} {2018})},\ \Eprint {https://arxiv.org/abs/1801.10122} {arXiv:1801.10122 [hep-lat]} \BibitemShut {NoStop}%
\bibitem [{\citenamefont {Achasov}\ and\ \citenamefont {Shestakov}(2025)}]{Achasov:2025hkv}%
  \BibitemOpen
  \bibfield  {author} {\bibinfo {author} {\bibfnamefont {N.~N.}\ \bibnamefont {Achasov}}\ and\ \bibinfo {author} {\bibfnamefont {G.~N.}\ \bibnamefont {Shestakov}},\ }\bibinfo {title} {{Phenomenological description of the $D_{s0}^{*}(2317) \textrightarrow D_{s}\pi^{0}$ decay}},\ \href {https://doi.org/10.1103/jwkb-3zl9} {\bibfield  {journal} {\bibinfo  {journal} {Phys. Rev. D}\ }\textbf {\bibinfo {volume} {112}},\ \bibinfo {pages} {096004} (\bibinfo {year} {2025})},\ \Eprint {https://arxiv.org/abs/2508.15224} {arXiv:2508.15224 [hep-ph]} \BibitemShut {NoStop}%
\bibitem [{\citenamefont {Yue}\ {\it et~al.}(2026)\citenamefont {Yue}, \citenamefont {Guo}, \citenamefont {Chen},\ and\ \citenamefont {Santopinto}}]{Yue:2025wcl}%
  \BibitemOpen
  \bibfield  {author} {\bibinfo {author} {\bibfnamefont {Z.-L.}\ \bibnamefont {Yue}}, \bibinfo {author} {\bibfnamefont {Q.-Y.}\ \bibnamefont {Guo}}, \bibinfo {author} {\bibfnamefont {D.-Y.}\ \bibnamefont {Chen}},\ and\ \bibinfo {author} {\bibfnamefont {E.}~\bibnamefont {Santopinto}},\ }\bibinfo {title} {{Determining the width of $D_{s0}^{*}(2317)$ by using $T_{c\bar{s}0}^{a}(2327)$ in a molecular frame}},\ \href {https://doi.org/10.1140/epjc/s10052-025-15248-w} {\bibfield  {journal} {\bibinfo  {journal} {Eur. Phys. J. C}\ }\textbf {\bibinfo {volume} {86}},\ \bibinfo {pages} {33} (\bibinfo {year} {2026})},\ \Eprint {https://arxiv.org/abs/2507.19641} {arXiv:2507.19641 [hep-ph]} \BibitemShut {NoStop}%
\bibitem [{\citenamefont {Achasov}\ {\it et~al.}(1979)\citenamefont {Achasov}, \citenamefont {Devyanin},\ and\ \citenamefont {Shestakov}}]{Achasov:1979xc}%
  \BibitemOpen
  \bibfield  {author} {\bibinfo {author} {\bibfnamefont {N.~N.}\ \bibnamefont {Achasov}}, \bibinfo {author} {\bibfnamefont {S.~A.}\ \bibnamefont {Devyanin}},\ and\ \bibinfo {author} {\bibfnamefont {G.~N.}\ \bibnamefont {Shestakov}},\ }\bibinfo {title} {{The $S^{*}$- $\delta^{0}$ Mixing as the Threshold Phenomenon}},\ \href {https://doi.org/10.1016/0370-2693(79)90488-X} {\bibfield  {journal} {\bibinfo  {journal} {Phys. Lett. B}\ }\textbf {\bibinfo {volume} {88}},\ \bibinfo {pages} {367} (\bibinfo {year} {1979})}\BibitemShut {NoStop}%
\bibitem [{\citenamefont {Bramon}\ {\it et~al.}(1992)\citenamefont {Bramon}, \citenamefont {Grau},\ and\ \citenamefont {Pancheri}}]{Bramon:1992kr}%
  \BibitemOpen
  \bibfield  {author} {\bibinfo {author} {\bibfnamefont {A.}~\bibnamefont {Bramon}}, \bibinfo {author} {\bibfnamefont {A.}~\bibnamefont {Grau}},\ and\ \bibinfo {author} {\bibfnamefont {G.}~\bibnamefont {Pancheri}},\ }\bibinfo {title} {{Intermediate vector meson contributions to $V^{0} \rightarrow P^{0} P^{0} \gamma$ decays}},\ \href {https://doi.org/10.1016/0370-2693(92)90041-2} {\bibfield  {journal} {\bibinfo  {journal} {Phys. Lett. B}\ }\textbf {\bibinfo {volume} {283}},\ \bibinfo {pages} {416} (\bibinfo {year} {1992})}\BibitemShut {NoStop}%
\bibitem [{\citenamefont {Gamermann}\ {\it et~al.}(2010)\citenamefont {Gamermann}, \citenamefont {Nieves}, \citenamefont {Oset},\ and\ \citenamefont {Ruiz~Arriola}}]{Gamermann:2009uq}%
  \BibitemOpen
  \bibfield  {author} {\bibinfo {author} {\bibfnamefont {D.}~\bibnamefont {Gamermann}}, \bibinfo {author} {\bibfnamefont {J.}~\bibnamefont {Nieves}}, \bibinfo {author} {\bibfnamefont {E.}~\bibnamefont {Oset}},\ and\ \bibinfo {author} {\bibfnamefont {E.}~\bibnamefont {Ruiz~Arriola}},\ }\bibinfo {title} {{Couplings in coupled channels versus wave functions: application to the X(3872) resonance}},\ \href {https://doi.org/10.1103/PhysRevD.81.014029} {\bibfield  {journal} {\bibinfo  {journal} {Phys. Rev. D}\ }\textbf {\bibinfo {volume} {81}},\ \bibinfo {pages} {014029} (\bibinfo {year} {2010})},\ \Eprint {https://arxiv.org/abs/0911.4407} {arXiv:0911.4407 [hep-ph]} \BibitemShut {NoStop}%
\bibitem [{\citenamefont {Song}\ {\it et~al.}(2022)\citenamefont {Song}, \citenamefont {Dai},\ and\ \citenamefont {Oset}}]{Song:2022yvz}%
  \BibitemOpen
  \bibfield  {author} {\bibinfo {author} {\bibfnamefont {J.}~\bibnamefont {Song}}, \bibinfo {author} {\bibfnamefont {L.~R.}\ \bibnamefont {Dai}},\ and\ \bibinfo {author} {\bibfnamefont {E.}~\bibnamefont {Oset}},\ }\bibinfo {title} {{How much is the compositeness of a bound state constrained by $a$ and $r_0$? The role of the interaction range}},\ \href {https://doi.org/10.1140/epja/s10050-022-00753-3} {\bibfield  {journal} {\bibinfo  {journal} {Eur. Phys. J. A}\ }\textbf {\bibinfo {volume} {58}},\ \bibinfo {pages} {133} (\bibinfo {year} {2022})},\ \Eprint {https://arxiv.org/abs/2201.04414} {arXiv:2201.04414 [hep-ph]} \BibitemShut {NoStop}%
\bibitem [{\citenamefont {Lu}\ {\it et~al.}(2021)\citenamefont {Lu}, \citenamefont {Liu}, \citenamefont {Shi},\ and\ \citenamefont {Geng}}]{Lu:2021irg}%
  \BibitemOpen
  \bibfield  {author} {\bibinfo {author} {\bibfnamefont {J.-X.}\ \bibnamefont {Lu}}, \bibinfo {author} {\bibfnamefont {M.-Z.}\ \bibnamefont {Liu}}, \bibinfo {author} {\bibfnamefont {R.-X.}\ \bibnamefont {Shi}},\ and\ \bibinfo {author} {\bibfnamefont {L.-S.}\ \bibnamefont {Geng}},\ }\bibinfo {title} {{Understanding $P_{cs}(4459)$ as a hadronic molecule in the $\Xi_{b}^{-} \rightarrow J/\psi \Lambda K^{-}$ decay}},\ \href {https://doi.org/10.1103/PhysRevD.104.034022} {\bibfield  {journal} {\bibinfo  {journal} {Phys. Rev. D}\ }\textbf {\bibinfo {volume} {104}},\ \bibinfo {pages} {034022} (\bibinfo {year} {2021})},\ \Eprint {https://arxiv.org/abs/2104.10303} {arXiv:2104.10303 [hep-ph]} \BibitemShut {NoStop}%
\bibitem [{\citenamefont {Sakai}\ {\it et~al.}(2018)\citenamefont {Sakai}, \citenamefont {Oset},\ and\ \citenamefont {Ramos}}]{Sakai:2017hpg}%
  \BibitemOpen
  \bibfield  {author} {\bibinfo {author} {\bibfnamefont {S.}~\bibnamefont {Sakai}}, \bibinfo {author} {\bibfnamefont {E.}~\bibnamefont {Oset}},\ and\ \bibinfo {author} {\bibfnamefont {A.}~\bibnamefont {Ramos}},\ }\bibinfo {title} {{Triangle singularities in $B^-\rightarrow K^-\pi^-D_{s0}^+$ and $B^-\rightarrow K^-\pi^-D_{s1}^+$}},\ \href {https://doi.org/10.1140/epja/i2018-12450-5} {\bibfield  {journal} {\bibinfo  {journal} {Eur. Phys. J. A}\ }\textbf {\bibinfo {volume} {54}},\ \bibinfo {pages} {10} (\bibinfo {year} {2018})},\ \Eprint {https://arxiv.org/abs/1705.03694} {arXiv:1705.03694 [hep-ph]} \BibitemShut {NoStop}%
\bibitem [{\citenamefont {Oset}\ {\it et~al.}(2003)\citenamefont {Oset}, \citenamefont {Pelaez},\ and\ \citenamefont {Roca}}]{Oset:2002sh}%
  \BibitemOpen
  \bibfield  {author} {\bibinfo {author} {\bibfnamefont {E.}~\bibnamefont {Oset}}, \bibinfo {author} {\bibfnamefont {J.~R.}\ \bibnamefont {Pelaez}},\ and\ \bibinfo {author} {\bibfnamefont {L.}~\bibnamefont {Roca}},\ }\bibinfo {title} {{$\eta \rightarrow \pi^0 \gamma \gamma$ decay within a chiral unitary approach}},\ \href {https://doi.org/10.1103/PhysRevD.67.073013} {\bibfield  {journal} {\bibinfo  {journal} {Phys. Rev. D}\ }\textbf {\bibinfo {volume} {67}},\ \bibinfo {pages} {073013} (\bibinfo {year} {2003})},\ \Eprint {https://arxiv.org/abs/hep-ph/0210282} {arXiv:hep-ph/0210282} \BibitemShut {NoStop}%
\bibitem [{\citenamefont {Ramos}\ {\it et~al.}(2013)\citenamefont {Ramos}, \citenamefont {Tolos}, \citenamefont {Molina},\ and\ \citenamefont {Oset}}]{Ramos:2013mda}%
  \BibitemOpen
  \bibfield  {author} {\bibinfo {author} {\bibfnamefont {A.}~\bibnamefont {Ramos}}, \bibinfo {author} {\bibfnamefont {L.}~\bibnamefont {Tolos}}, \bibinfo {author} {\bibfnamefont {R.}~\bibnamefont {Molina}},\ and\ \bibinfo {author} {\bibfnamefont {E.}~\bibnamefont {Oset}},\ }\bibinfo {title} {{The width of the $\omega$ meson in the nuclear medium}},\ \href {https://doi.org/10.1140/epja/i2013-13148-x} {\bibfield  {journal} {\bibinfo  {journal} {Eur. Phys. J. A}\ }\textbf {\bibinfo {volume} {49}},\ \bibinfo {pages} {148} (\bibinfo {year} {2013})},\ \Eprint {https://arxiv.org/abs/1306.5921} {arXiv:1306.5921 [nucl-th]} \BibitemShut {NoStop}%
\bibitem [{\citenamefont {Dias}\ {\it et~al.}(2025)\citenamefont {Dias}, \citenamefont {Li},\ and\ \citenamefont {Oset}}]{Dias:2025izv}%
  \BibitemOpen
  \bibfield  {author} {\bibinfo {author} {\bibfnamefont {J.~M.}\ \bibnamefont {Dias}}, \bibinfo {author} {\bibfnamefont {Y.-Y.}\ \bibnamefont {Li}},\ and\ \bibinfo {author} {\bibfnamefont {E.}~\bibnamefont {Oset}},\ }\bibinfo {title} {{$\pi^+\pi^-$ and $D^+_s\pi^{\pm}$ mass spectra in the $D_{s1}(2536) \to D^+_s\pi^+\pi^-$ decays}},\ \href {https://doi.org/10.1103/nr6f-cpb7} {\bibfield  {journal} {\bibinfo  {journal} {Phys. Rev. D}\ }\textbf {\bibinfo {volume} {112}},\ \bibinfo {pages} {114018} (\bibinfo {year} {2025})},\ \Eprint {https://arxiv.org/abs/2507.20425} {arXiv:2507.20425 [hep-ph]} \BibitemShut {NoStop}%
\bibitem [{\citenamefont {Cho}\ and\ \citenamefont {Wise}(1994)}]{Cho:1994zu}%
  \BibitemOpen
  \bibfield  {author} {\bibinfo {author} {\bibfnamefont {P.~L.}\ \bibnamefont {Cho}}\ and\ \bibinfo {author} {\bibfnamefont {M.~B.}\ \bibnamefont {Wise}},\ }\bibinfo {title} {{Remarks on $D^*_s \to D_s \pi^0$ decay}},\ \href {https://doi.org/10.1103/PhysRevD.49.6228} {\bibfield  {journal} {\bibinfo  {journal} {Phys. Rev. D}\ }\textbf {\bibinfo {volume} {49}},\ \bibinfo {pages} {6228} (\bibinfo {year} {1994})},\ \Eprint {https://arxiv.org/abs/hep-ph/9401301} {arXiv:hep-ph/9401301} \BibitemShut {NoStop}%
\bibitem [{\citenamefont {Godfrey}\ and\ \citenamefont {Isgur}(1985)}]{Godfrey:1985xj}%
  \BibitemOpen
  \bibfield  {author} {\bibinfo {author} {\bibfnamefont {S.}~\bibnamefont {Godfrey}}\ and\ \bibinfo {author} {\bibfnamefont {N.}~\bibnamefont {Isgur}},\ }\bibinfo {title} {{Mesons in a Relativized Quark Model with Chromodynamics}},\ \href {https://doi.org/10.1103/PhysRevD.32.189} {\bibfield  {journal} {\bibinfo  {journal} {Phys. Rev. D}\ }\textbf {\bibinfo {volume} {32}},\ \bibinfo {pages} {189} (\bibinfo {year} {1985})}\BibitemShut {NoStop}%
\bibitem [{\citenamefont {Ebert}\ {\it et~al.}(1998)\citenamefont {Ebert}, \citenamefont {Galkin},\ and\ \citenamefont {Faustov}}]{Ebert:1997nk}%
  \BibitemOpen
  \bibfield  {author} {\bibinfo {author} {\bibfnamefont {D.}~\bibnamefont {Ebert}}, \bibinfo {author} {\bibfnamefont {V.~O.}\ \bibnamefont {Galkin}},\ and\ \bibinfo {author} {\bibfnamefont {R.~N.}\ \bibnamefont {Faustov}},\ }\bibinfo {title} {{Mass spectrum of orbitally and radially excited heavy - light mesons in the relativistic quark model}},\ \href {https://doi.org/10.1103/PhysRevD.59.019902} {\bibfield  {journal} {\bibinfo  {journal} {Phys. Rev. D}\ }\textbf {\bibinfo {volume} {57}},\ \bibinfo {pages} {5663} (\bibinfo {year} {1998})},\ \bibinfo {note} {[Erratum: Phys.Rev.D 59, 019902 (1999)]},\ \Eprint {https://arxiv.org/abs/hep-ph/9712318} {arXiv:hep-ph/9712318} \BibitemShut {NoStop}%
\bibitem [{\citenamefont {Wei}\ {\it et~al.}(2006)\citenamefont {Wei}, \citenamefont {Huang},\ and\ \citenamefont {Zhu}}]{Wei:2005ag}%
  \BibitemOpen
  \bibfield  {author} {\bibinfo {author} {\bibfnamefont {W.}~\bibnamefont {Wei}}, \bibinfo {author} {\bibfnamefont {P.-Z.}\ \bibnamefont {Huang}},\ and\ \bibinfo {author} {\bibfnamefont {S.-L.}\ \bibnamefont {Zhu}},\ }\bibinfo {title} {{Strong decays of $D_{sJ}(2317)$ and $D_{sJ}(2460)$}},\ \href {https://doi.org/10.1103/PhysRevD.73.034004} {\bibfield  {journal} {\bibinfo  {journal} {Phys. Rev. D}\ }\textbf {\bibinfo {volume} {73}},\ \bibinfo {pages} {034004} (\bibinfo {year} {2006})},\ \Eprint {https://arxiv.org/abs/hep-ph/0510039} {arXiv:hep-ph/0510039} \BibitemShut {NoStop}%
\bibitem [{\citenamefont {Goity}\ and\ \citenamefont {Roberts}(2001)}]{Goity:2000dk}%
  \BibitemOpen
  \bibfield  {author} {\bibinfo {author} {\bibfnamefont {J.~L.}\ \bibnamefont {Goity}}\ and\ \bibinfo {author} {\bibfnamefont {W.}~\bibnamefont {Roberts}},\ }\bibinfo {title} {{Radiative Transitions in Heavy Mesons in a Relativistic Quark Model}},\ \href {https://doi.org/10.1103/PhysRevD.64.094007} {\bibfield  {journal} {\bibinfo  {journal} {Phys. Rev. D}\ }\textbf {\bibinfo {volume} {64}},\ \bibinfo {pages} {094007} (\bibinfo {year} {2001})},\ \Eprint {https://arxiv.org/abs/hep-ph/0012314} {arXiv:hep-ph/0012314} \BibitemShut {NoStop}%
\bibitem [{\citenamefont {Lu}\ {\it et~al.}(2006)\citenamefont {Lu}, \citenamefont {Chen}, \citenamefont {Deng},\ and\ \citenamefont {Zhu}}]{Lu:2006ry}%
  \BibitemOpen
  \bibfield  {author} {\bibinfo {author} {\bibfnamefont {J.}~\bibnamefont {Lu}}, \bibinfo {author} {\bibfnamefont {X.-L.}\ \bibnamefont {Chen}}, \bibinfo {author} {\bibfnamefont {W.-Z.}\ \bibnamefont {Deng}},\ and\ \bibinfo {author} {\bibfnamefont {S.-L.}\ \bibnamefont {Zhu}},\ }\bibinfo {title} {{Pionic decays of $D_{sj}(2317)$, $D_{sj}(2460)$ and $B_{sj}(5718)$, $B_{sj}(5765)$}},\ \href {https://doi.org/10.1103/PhysRevD.73.054012} {\bibfield  {journal} {\bibinfo  {journal} {Phys. Rev. D}\ }\textbf {\bibinfo {volume} {73}},\ \bibinfo {pages} {054012} (\bibinfo {year} {2006})},\ \Eprint {https://arxiv.org/abs/hep-ph/0602167} {arXiv:hep-ph/0602167} \BibitemShut {NoStop}%
\bibitem [{\citenamefont {Wang}\ {\it et~al.}(2006)\citenamefont {Wang}, \citenamefont {Chen}, \citenamefont {Lu}, \citenamefont {Zhu},\ and\ \citenamefont {Deng}}]{Wang:2006fg}%
  \BibitemOpen
  \bibfield  {author} {\bibinfo {author} {\bibfnamefont {F.-L.}\ \bibnamefont {Wang}}, \bibinfo {author} {\bibfnamefont {X.-L.}\ \bibnamefont {Chen}}, \bibinfo {author} {\bibfnamefont {D.-H.}\ \bibnamefont {Lu}}, \bibinfo {author} {\bibfnamefont {S.-L.}\ \bibnamefont {Zhu}},\ and\ \bibinfo {author} {\bibfnamefont {W.-Z.}\ \bibnamefont {Deng}},\ }\bibinfo {title} {{Decays of $D^*_{sj}(2317)$ and $D_{sj}(2460)$ Mesons in the Quark Model}},\ \href@noop {} {\bibfield  {journal} {\bibinfo  {journal} {HEPNP}\ }\textbf {\bibinfo {volume} {30}},\ \bibinfo {pages} {1041} (\bibinfo {year} {2006})},\ \Eprint {https://arxiv.org/abs/hep-ph/0604090} {arXiv:hep-ph/0604090} \BibitemShut {NoStop}%
\bibitem [{\citenamefont {Yang}\ {\it et~al.}(2020)\citenamefont {Yang}, \citenamefont {Wang}, \citenamefont {Meng},\ and\ \citenamefont {Zhu}}]{Yang:2019cat}%
  \BibitemOpen
  \bibfield  {author} {\bibinfo {author} {\bibfnamefont {B.}~\bibnamefont {Yang}}, \bibinfo {author} {\bibfnamefont {B.}~\bibnamefont {Wang}}, \bibinfo {author} {\bibfnamefont {L.}~\bibnamefont {Meng}},\ and\ \bibinfo {author} {\bibfnamefont {S.-L.}\ \bibnamefont {Zhu}},\ }\bibinfo {title} {{Isospin violating decay $D^*_s \to D_s \pi^0$ in chiral perturbation theory}},\ \href {https://doi.org/10.1103/PhysRevD.101.054019} {\bibfield  {journal} {\bibinfo  {journal} {Phys. Rev. D}\ }\textbf {\bibinfo {volume} {101}},\ \bibinfo {pages} {054019} (\bibinfo {year} {2020})},\ \Eprint {https://arxiv.org/abs/1912.09616} {arXiv:1912.09616 [hep-ph]} \BibitemShut {NoStop}%
\bibitem [{\citenamefont {Fu}\ {\it et~al.}(2025)\citenamefont {Fu}, \citenamefont {Guo}, \citenamefont {Hanhart},\ and\ \citenamefont {Nefediev}}]{Fu:2025lfo}%
  \BibitemOpen
  \bibfield  {author} {\bibinfo {author} {\bibfnamefont {H.-L.}\ \bibnamefont {Fu}}, \bibinfo {author} {\bibfnamefont {F.-K.}\ \bibnamefont {Guo}}, \bibinfo {author} {\bibfnamefont {C.}~\bibnamefont {Hanhart}},\ and\ \bibinfo {author} {\bibfnamefont {A.}~\bibnamefont {Nefediev}},\ }\href@noop {} {\bibinfo {title} {{What can we learn from the radiative decays of the $D_{s1}(2460)$ meson?}}} (\bibinfo {year} {2025}),\ \Eprint {https://arxiv.org/abs/2512.05476} {arXiv:2512.05476 [hep-ph]} \BibitemShut {NoStop}%
\bibitem [{\citenamefont {Navas}\ {\it et~al.}(2024)\citenamefont {Navas} {\it et~al.}}]{ParticleDataGroup:2024cfk}%
  \BibitemOpen
  \bibfield  {author} {\bibinfo {author} {\bibfnamefont {S.}~\bibnamefont {Navas}} {\it et~al.} (\bibinfo {collaboration} {Particle Data Group}),\ }\bibinfo {title} {{Review of particle physics}},\ \href {https://doi.org/10.1103/PhysRevD.110.030001} {\bibfield  {journal} {\bibinfo  {journal} {Phys. Rev. D}\ }\textbf {\bibinfo {volume} {110}},\ \bibinfo {pages} {030001} (\bibinfo {year} {2024})}\BibitemShut {NoStop}%
\bibitem [{\citenamefont {Besson}\ {\it et~al.}(2003)\citenamefont {Besson} {\it et~al.}}]{CLEO:2003ggt}%
  \BibitemOpen
  \bibfield  {author} {\bibinfo {author} {\bibfnamefont {D.}~\bibnamefont {Besson}} {\it et~al.} (\bibinfo {collaboration} {CLEO}),\ }\bibinfo {title} {{Observation of a narrow resonance of mass 2.46 GeV/$c^2$ decaying to $D_s^{*+} \pi^0$ and confirmation of the $D^*_{sJ}(2317)$ state}},\ \href {https://doi.org/10.1103/PhysRevD.68.032002} {\bibfield  {journal} {\bibinfo  {journal} {Phys. Rev. D}\ }\textbf {\bibinfo {volume} {68}},\ \bibinfo {pages} {032002} (\bibinfo {year} {2003})},\ \bibinfo {note} {[Erratum: Phys.Rev.D 75, 119908 (2007)]},\ \Eprint {https://arxiv.org/abs/hep-ex/0305100} {arXiv:hep-ex/0305100} \BibitemShut {NoStop}%
\end{thebibliography}%
\end{document}